\documentclass[12pt,a4,sans]{article}
\pdfoutput=1
\usepackage[latin1]{inputenc}
\usepackage{amssymb}
\usepackage{amsmath}
\usepackage{amsthm}
\usepackage{amsfonts}
\usepackage[pdftex]{graphicx}
\usepackage{geometry}
\usepackage{multirow}
\usepackage{braket}
\usepackage{enumerate}

\allowdisplaybreaks

\usepackage[table,usenames,dvipsnames]{xcolor}

\usepackage{setspace}
\usepackage[backgroundcolor=white,linecolor=black]{todonotes}
\usepackage[in]{fullpage}
\usepackage{hyperref}
\usepackage{array}
\usepackage{cancel}
\usepackage{wrapfig}
\usepackage[font=small,labelfont=bf]{caption}
\usepackage{pifont}
\newcommand{\cmark}{\ding{51}}
\newcommand{\xmark}{\ding{55}}
\usepackage[dvipsnames]{xcolor}
\usepackage{mathtools}

\usepackage{color}

\numberwithin{equation}{section}

\hypersetup{
	colorlinks=true,
	linktoc=page,
	citecolor=Blue,
	linkcolor=Blue,
	urlcolor=Blue} 

\makeatletter 
\renewcommand{\maketitle} 
{ \begingroup \begin{center} \large {\bf \@title}
		\vskip 5pt \large \@author \\ \vskip 5pt \@date \end{center}
	\vskip 5pt \endgroup \setcounter{footnote}{0} }
\makeatother 

\newcommand{\pic}[2]{\vcenter{\hbox{\includegraphics[scale=#1]{#2}}}}

\newcommand{\comments}[1]{}
\newcommand{\la}{\langle}
\newcommand{\ra}{\rangle}

\newcommand{\N}{\mathcal{N}}

\newcommand{\Tr}{\text{Tr}}

\newcommand{\OC}{\mathcal{O}_{\mathcal{C}}}
\newcommand{\OS}{\mathcal{O}_{\mathcal{S}}}

\renewcommand{\O}{\mathcal{O}}

\newcommand{\be}{\begin{equation}}
\newcommand{\ee}{\end{equation}}

\newcommand{\hf}{\frac{1}{2}}

\def\beqa{\begin{eqnarray}}
\def\eeqa{\end{eqnarray}}
\def\beq{\begin{equation}}
\def\eeq{\end{equation}}

\def\Tr{{\rm Tr}}
\def\one{\mbox{1 \kern-.59em {\rm l}}}

 \def\C_B{{\cal B}} \def\cC{{\cal C}}
  \def\C_F{{\cal F}}
  
  \def\cL{{\cal L}}
\def\cM{{\cal M}} \def\cN{{\cal N}} \def\cO{{\cal O}}
  \def\cR{{\cal R}}
\def\cS{{\cal S}}

\def\uno{\mbox{1 \kern-.59em {\rm l}}}

\def\one{1\!\!1\,\,}

\def\bcomment#1{}

\def\eps{\epsilon}

\usepackage{slashed}
\usepackage{caption}

\usepackage[nottoc,notlot,notlof]{tocbibind}
\usepackage[nosort]{cite}
\usepackage{color}

\usepackage{parskip}

\graphicspath{{./Images/}}

\long\def\symbolfootnote[#1]#2{\begingroup%
	\def\thefootnote{\fnsymbol{footnote}}\footnote[#1]{#2}\endgroup}

\begin{document}
	
	\begin{flushright}
		QMUL-PH-18-05\\
		CERN-TH-2018-064
	\end{flushright}
	
	\vspace{15pt} 

	\begin{center}
		
		{\Large \bf  $\Tr(F^3)$ supersymmetric form  factors }\\ 
		\vspace{0.3 cm} {\Large \bf   
		and maximal transcendentality}\\ 
                  \vspace{0.3 cm}
		{\Large \bf  Part II:  $0\!<\!\cN\!<\!4$ super Yang-Mills}

		%
		\vspace{25pt}

		{\mbox {\bf  Andreas~Brandhuber$^{a,\S}$, 
				Martyna~Kostaci\'{n}ska$^{a,\S}$,}} \\ \vspace{0.2cm}
				{\mbox{\bf
				Brenda~Penante$^{b,\star}$ and  %
				Gabriele~Travaglini$^{a,\S}$}}%

		\vspace{0.5cm}
		
		\begin{quote}
			{\small \em
				\begin{itemize}
					\item[\ \ \ \ \ \ $^a$]
					\begin{flushleft}
						Centre for Research in String Theory\\
						School of Physics and Astronomy\\
						Queen Mary University of London\\
						Mile End Road, London E1 4NS, United Kingdom
					\end{flushleft}

					\item[\ \ \ \ \ \ $^b$]CERN Theory Division, 1211 Geneva 23, Switzerland

				\end{itemize}
			}
		\end{quote}


		\vspace{15pt}  

{\bf Abstract}
	\end{center}
	
	\vspace{0.3cm} 
	
\noindent

\noindent
The study of form factors has many phenomenologically interesting applications, one of which is  Higgs plus gluon amplitudes in QCD. Through  effective field theory techniques these are related to form factors of various operators of increasing classical dimension.   In this paper we extend  our analysis of the  first finite top-mass  correction, arising from the operator ${\rm Tr}(F^3)$,  from $\cN\!=\!4$ super Yang-Mills to theories with $\cN\!<\!4$, for the case of three gluons and up to two loops.  
We confirm our earlier result that  the maximally transcendental part of the associated Catani remainder is universal and equal to that of the form factor of a protected trilinear operator in the maximally supersymmetric theory. The terms with lower transcendentality deviate from the $\cN\!=\!4$ answer by a surprisingly small set of terms involving for example $\zeta_2$,  $\zeta_3$ and simple powers of logarithms, for which we provide explicit expressions.

	\vfill
	\hrulefill
	\newline
\vspace{-1cm}
$^{\S}$~\!\!{\tt\footnotesize\{a.brandhuber, m.m.kostacinska, g.travaglini\}@qmul.ac.uk}, \ $^{\star}$~\!\!{\tt\footnotesize b.penante@cern.ch}

	\setcounter{page}{0}
	\thispagestyle{empty}
	\newpage


	\setcounter{tocdepth}{4}
	\hrule height 0.75pt
	\tableofcontents
	\vspace{0.8cm}
	\hrule height 0.75pt
	\vspace{1cm}
	
	\setcounter{tocdepth}{2}

	\newpage

\section{Introduction}\label{sec:Introduction}

In this paper we extend the study of  form factors of the operator ${\rm Tr} (F^3)$ initiated in \cite{Brandhuber:2017bkg, Part1} at two loops  with an external state containing three gluons of positive helicity. The importance of these form factors arises from their connection to the effective theory for Higgs plus many gluon processes. In this approach, the one-loop gluon-fusion diagram involving a loop of top quarks is replaced by a set of local interactions in an expansion in $1/m_t$ where $m_t$ is the top mass. This expansion has the form~\cite{Neill:2009tn,Dawson:2014ora}
\begin{align}
\label{one}
\cL_{\rm eff} \, = \, \hat{C}_0 \O_0 \, + \, {1\over m_{t}^2} \sum_{i=1}^4 \hat{C}_i \O_i \, + \, \cO
\left({1\over m_{t}^4} \right)\ , 
\end{align}
where $\O_i$, $i=1, \ldots , 4$ are dimension-7 operators made of gluon field strengths and covariant derivatives, and  $\O_0 \coloneqq H \, \Tr (F^2)$.  
$\hat{C}_0$, $\hat{C}_i$ are the  matching coefficients and are proportional to $1/v$, where $v$ is the Higgs field  vacuum expectation value.
Due to the equations of motion, in pure Yang-Mills one can eliminate two of the four operators in the sum   
\cite{Dawson:2014ora,Gracey:2002he}, 
and the remaining two operators can be chosen to be $H\, \Tr (F^3)$ and $H\, \Tr \big(D^\mu F^{\nu \rho}D_\mu F_{\nu \rho}\big)$.
One is then led to the study of the form factors of the two operators 
\beq
\Tr (F^3) =    \Tr (F_{\rm ASD} ^3) + \Tr (F_{\rm SD} ^3) \propto \cO_{\cC} + \overline{\cO}_{\cC} \ , \qquad   \O_\cM \,\propto  \, \Tr \big(D^\mu F^{\nu \rho}D_\mu F_{\nu \rho}\big)
\ , 
\eeq
where ASD  stands for the anti-selfdual part of the gluon field strength (which is the only part contributing at two loops for our external state). 

Our main goal is to   identify some universal structures in the expressions of such form factors, in particular across different classes of operators and for various amounts of supersymmetry. Several hints of this universality have  already been found in related investigations. In particular, in \cite{Brandhuber:2012vm} it was found that the form factor remainder for the half-BPS bilinear scalar  operator ${\rm Tr}(X^2)$ in  $\cN\!=\!4$ super Yang-Mills (SYM) captures the maximally transcendental part of the  remainder computed in pure Yang-Mills of the operator ${\rm Tr} (F^2)$ with a state of three gluons \cite{Gehrmann:2011aa}.%
\footnote{Here $X$ denotes one of the three complex scalar fields of the $\cN\!=\!4$ theory.} 
In turn, these particular form factors compute the leading-order Higgs plus gluon amplitudes in the   $1/m_t$ expansion, related to the term $\cO_0$ in~\eqref{one}.

This surprising coincidence was the motivation for the study begun in \cite{Penante:2014sza,Brandhuber:2014ica,Brandhuber:2016fni} of form factors  of operators containing three scalar fields  in $\cN\!=\!4$ SYM. In particular, it was found in \cite{Brandhuber:2016fni} that at two loops, the minimal form factor for the non-protected operator $ {\rm Tr}  (X[Y, Z])$ has the same maximally  transcendental part of the minimal form factor remainder of the protected operator ${\rm Tr}  (X^3)$. 
The $ {\rm Tr}  (X[Y, Z])$  operator (or more precisely a certain admixture of it with a fermion bilinear) is a descendant of the simplest non-protected operator, namely the Konishi. 
While the form of universality we alluded to earlier is across different {\it theories}, this new appearance  is across different types of {\it operators}. 
Other purely transcendental terms of decreasing transcendentality three to zero (which we will refer to as  ``pure" terms)  were found  in the remainder for ${\rm Tr}  (X[Y, Z])$, and unexpected connections of these terms to certain spin-chain remainder  densities in the $SU(2)$ sector  \cite{Loebbert:2015ova} were identified. 
This was quite surprising since the operator $ {\rm Tr} (X[Y, Z])$ belongs to a different sector, namely the $SU(2|3)$ sector \cite{Beisert:2003ys}.

The  calculation of \cite{Brandhuber:2016fni} was a stepping stone for the computations of the form factors  of the operator $\cO_\cC$ in $\cN\!=\!4$ SYM  in \cite{Brandhuber:2017bkg}. More precisely, in that paper two different operators were considered: $\cO_\cC$ and a particular supersymmetric completion thereof denoted by $\cO_\cS$, belonging to the Konishi supermultiplet, whose MHV form factors have recently been computed \cite{Chicherin:2016qsf}. It was found in \cite{Brandhuber:2017bkg} that the maximally transcendental part of these form factors with an external state of three gluons is one and the same across theories with {\it any} amount of supersymmetry, including pure Yang-Mills (or QCD), and also identical for $\cO_\cS$ and $\cO_\cC$. These form factors in turn describe the first subleading corrections to Higgs + many gluons in the $1/m_t$ expansion. Unlike the case of the operator $ {\rm Tr} (X[Y, Z])$, the remainders for $\cO_\cS$ and $\cO_\cC$ show a new feature in that they are  accompanied by ``non-pure" terms, {\it i.e.} terms of transcendentality degree ranging from three to zero which are further multiplied by ratios of kinematic invariants. Interesting relations across terms with varying degree of transcendentality were observed  in \cite{Part1} as a consequence of requiring the absence of unphysical singularities in soft/collinear limits.

In this paper we quantify these findings  by providing   explicit expressions for the remainder functions in $\cN\!=\!2,1$ SYM, both for the component operator $\cO_\cC$ and for its supersymmetric version $\cO_\cS$, whose form factors can be simply obtained by a truncation \cite{Elvang:2011fx} of the result of  \cite{Chicherin:2016qsf} (we note in passing that we will never need to know the explicit expression of the operator $\cO_\cS$, only of its MHV super form factors).

An important  disclaimer is in order here. Throughout our calculations we  use  four-dimensional  expressions of amplitudes and form factors as input in the unitarity cuts. 
As mentioned in \cite{Part1}, 
there are examples in $\cN\!=\!4$ SYM where it has explicitly been observed that  four-dimensional cuts are sufficient for computing finite remainders, namely  for  four- \cite{0309040},  five-  \cite{Bern:2006vw} and six-point   \cite{Bern:2008ap}  two-loop  remainders of MHV amplitudes. This happens because of the absence of so-called $\mu^2$-terms (that can only be detected  by performing cuts in $D$ dimensions) at four points, and because of remarkable cancellations  in the five- and six-point cases which occur thanks to the particular definition of the remainder function. 
To the best of our knowledge, no such examples exist with $\cN\!<\!4$ supersymmetry. We cannot a priori exclude the presence of such  $\mu^2$-terms, and the potential modifications to the finite remainder function they could induce, however we do mention that 
our result passes several   consistency checks. These include  reproducing the correct infrared and ultraviolet divergences, and soft/collinear factorisation at two loops. 
Furthermore, we observe  that  the relevant one-loop form factor used throughout this paper as obtained from  four-dimensional cuts is also  correct in  $D$ dimensions \cite{Neill:2009mz}, {\it i.e.}~its expression has no additional $\mu^2$-terms. This quantity plays a twofold r\^{o}le, in that it enters cuts of two-loop form factors, and  is also used in the definition of our two-loop remainders.

The results of our investigation can be summarised as follows: 

\begin{itemize}
\item[{\bf 1.}] The maximally transcendental part of the form factors of the operators $\cO_\cS$ and $\cO_\cC$ is the same as that of the half-BPS operator ${\rm Tr} (X^3)$ in the $\cN\!=\!4$ SYM theory, regardless of the amount of supersymmetry (including 
 $\cN\!=\!0$) \cite{Brandhuber:2017bkg}. The latter statement was confirmed by a recent explicit computation in \cite{Jin:2018fak}.

\item[{\bf 2.}] The non-pure terms of our remainders are identical to those computed in the maximally supersymmetric theory. 

\item[{\bf 3.}] The only differences arise in the pure terms at transcendentality below four, and are limited to a very restricted type of terms involving   $\zeta_2$,  $\zeta_3$ and simple powers of logarithms (after disentangling the mixing). The results of our calculations are collected in 
Tables \ref{tab:difference-remainders} and \ref{tab:difference-remainders2}. 

\end{itemize}

The rest of the paper is organised as follows. In Section \ref{Sec:2} we briefly discuss the operators studied in this paper and their tree-level form factors, while  in Section \ref{Sec:3} we summarise the one-loop calculation. In Section \ref{Sec:4} we move on to calculate the two-loop minimal form factors in theories with less than maximal supersymmetry. In Section \ref{Sec:7} we compute the Catani two-loop form factor remainder functions in $\N\!=\!2,1$ SYM. 
We conclude in Section~\ref{Sec:Discussion} with a discussion of our results, their implications, and a number of  consistency checks.

\section{Operators and tree-level form factors in $\cN=1,2,4$}\label{Sec:2}

As explained in detail in \cite{Brandhuber:2017bkg,Part1}, a central point of our discussion consists of appropriately translating the operator 
 $\O_\mathcal{C}\propto\Tr(F^3_{\mathrm{ASD}})$ to a supersymmetric completion  
 $\O_{\mathcal  S}\!= \!  \O_{\mathcal  C}+\O(g)$. In \cite{Brandhuber:2017bkg} we have identified $\O_{\mathcal  S}$ for the case of $\N\!=\!4$  SYM as 
 a $\cS$upersymmetric descendant of the Konishi, generated by acting with tree-level supercharges on the lowest-dimensional operator in the multiplet. Notably,  the $\cC$omponent operator $\O_\cC$ is   contained within $\O_{\mathcal  S}$. 
 
 The key point to make here is that 
 similar supersymmetric completions of $\O_\cC$ can be obtained in $\N\!=\!2, 1$ SYM by an appropriate truncation  \cite{Elvang:2011fx}. We will see shortly that for the concrete calculations in this paper, we will only need $\O_{\mathcal  S}$ for $\N\!=\!2$ SYM.

 We now review some of the ingredients of the calculations. For both operators, the   tree-level minimal form factor with the external state of three positive-helicity gluons is given~by 
 \begin{align}
\label{eq:tree-level-F3}
F^{(0)}_{\O_\cS,\O_\cC}(1^+,2^+,3^+;q) \,=\,  -[12][23][31] \,.
\end{align}
Next, we recall the tree-level MHV super form factors \cite{Brandhuber:2011tv} 
of the full Konishi multiplet in $\cN \!=\! 4$ SYM have been constructed  and  expressed in a compact formula in \cite{Chicherin:2016qsf},
\begin{align}\label{eq:FF-Konishi}
\begin{split}
\hspace{-0.15cm}&\langle 1, 2, \ldots , n | {\mathcal K} (\theta, \bar\theta) | 0 \rangle_{\rm MHV}^{(0)}= 
{ {e}^{\sum_{l=1}^n [ l |\bar\theta\theta |l\rangle + \eta_l \langle \theta l \rangle } \over \langle 1 2\rangle \cdots \langle n  1 \rangle }\sum_{i\leq j < k \leq l}\!\!
( 2 \!-\! \delta_{ij}) ( 2 \!-\! \delta_{kl})
\epsilon^{ABCD} \hat\eta_{iA}\hat\eta_{jB}\hat\eta_{kC}\hat\eta_{lD}\langle jk\rangle \langle li \rangle \,, 
\end{split}
\end{align}
where $\hat{\eta}_A\coloneqq \eta_A + 2 [ \tilde\lambda \, \bar\theta_A]$ and $\eta_A$ are the usual on-shell superspace coordinates labelling the external on-shell states. The $\theta^A_\alpha$ and $\bar{\theta}_{A \dot\alpha}$ label the components of the Konishi super-multiplet. MHV form factors of $\O_{\mathcal K}$ are obtained by setting $\theta = \bar{\theta}=0$, while the form factors of $\overline{\O}_{\mathcal S}$ are obtained by setting $\bar{\theta}=0$
and extracting the $\theta^8$-term:
\begin{align}\label{eq:FF-OSbar}
\begin{split}
F^{(0)}_{\overline{\O}_{\mathcal S}, \mathrm{MHV}} (1, 2, \ldots , n ;q) =  
\frac{1}{144}{ \delta^{(8)}(\sum_{i=1}^n \eta_i \lambda_i ) \over \langle 1 2\rangle \cdots \langle n  1 \rangle }\sum_{i\leq j < k \leq l}\!\!
( 2 \!-\! \delta_{ij}) ( 2 \!-\! \delta_{kl})
\epsilon^{ABCD} \eta_{iA}\eta_{jB}\eta_{kC}\eta_{lD}\langle jk\rangle \langle li \rangle \, .
\end{split}
\end{align}
More details on the form of the operator $\mathcal{O}_{\mathcal S}$ can be found in Section
2.2 of \cite{Part1} and in particular a number of examples of four-point tree-level form factors relevant to unitarity cuts below are given in (2.13)-(2.20) of \cite{Part1}, describing the differences between
$\mathcal{O}_{\mathcal S}$ and~$\mathcal{O}_{\mathcal C}$.

{\bf Truncation to $\cN\!=\!2$ and  $\cN\!=\!1$ SYM}. Following \cite{Elvang:2011fx}, we can truncate formula \eqref{eq:FF-OSbar} to find the corresponding quantity in  $\N\!=\!2$ SYM. This will contain the operator ${\rm Tr}  (F^3)$, with appropriate additional $\cN\!=\!2$ completion terms. 
In order to do so, we first recall the form of the Nair on-shell superfields for $\N\!=\!4$, $\N\!=\!2$ and $\N\!=\!1$ SYM.  These are given by: 
\begin{align}
\begin{split}
\N\!=\!4: &   \qquad g^{(+)} (p) +  \psi^{A}(p)\eta_A  + {1\over 2}  \phi^{AB}(p)
\eta_A \eta_B +   {1\over 3!} \bar{\psi}^{ABC}(p)  \eta_A \eta_B\eta_C 
 + g^{(-)}(p)  \eta_1\cdots \eta_4 \ , 
 \\
 \N\!=\!2: &   \qquad g^{(+)} (p) + \sum_{I=1}^2  \psi^{I}(p)  \eta_I+ S
\eta_1 \eta_2 +   \Big( \bar{S} +   \sum_{I=1}^2 \bar{\psi}^{I34}(p)  \, \eta_I 
 + g^{(-)}(p)  \eta_1 \eta_2 \Big)\eta_3 \eta_4 \ , 
  \\
 \N\!=\!1: &   \qquad g^{(+)} (p) + \psi^1(p) \,  \eta_1  +   \Big(  \bar{\psi}^{234}(p)  
 + g^{(-)}(p)  \eta_1  \Big) \eta_2 \eta_3 \eta_4 \ ,
\end{split}
\end{align}
where in the first line $A,B,C=1, \ldots, 4$. 

In order to reduce \eqref{eq:FF-OSbar} to the form appropriate for
$\N\!=\!2$ SYM we have to project the superfields for each external particle. In practice this means that we drop all
terms which are linear in $\eta_3$ or $\eta_4$ for each particle in an  $\N\!=\!4$ super form factor and super amplitude.
The state sums in unitarity cuts are still performed using $\int\! d^4\eta$ for each internal leg.

We can apply the same procedure to the case of $\N\!=\!1$ SYM, however the supersymmetric completion of ${\rm Tr} \, (F^3)$ would only introduce
additional four-gluino terms which at our perturbative order and with our external state cannot contribute and hence are dropped.

\section{One-loop minimal form factors}\label{Sec:3}

For the reader's convenience we quote here the one-loop correction to the minimal form factor of the operators 
$\O_{\mathcal  S}$ and $\O_{\mathcal  C}$, calculated in \cite{Neill:2009mz,Part1}\footnote{Expressions for the one-loop master integrals can be found in Appendix \ref{App:Integrals}.}: 
\begin{align}\label{eq:one-loop-result}
F^{(1)}_{\O_{\mathcal  S},\O_{\mathcal  C}}(1^+,2^+,3^+;q)\,&=\,i\,F^{(0)}_{\O_{\mathcal  S},\O_{\mathcal  C}} \left(2\times\pic{0.4}{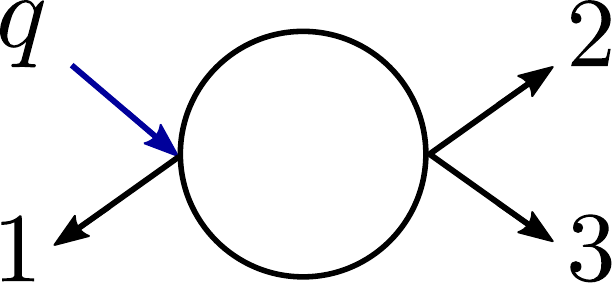}\,+\,s_{23}\times \pic{0.4}{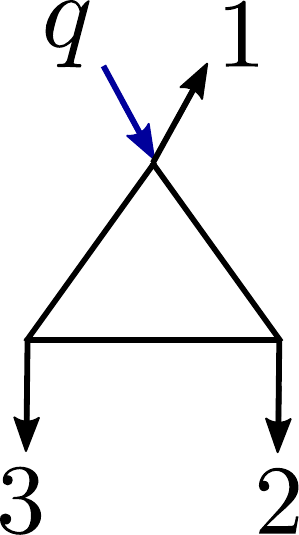}\,+\,\text{cyclic}(1,2,3)\right)\,.
\end{align}
For the purpose of the current discussion an important observation is in order here. The result for the one-loop form factor of the two operators $\O_\cC$ and $\O_\cS$ is not only operator-independent, but also theory-independent, {\it i.e.} the same whether computed in pure or supersymmetric Yang-Mills. This is due to the fact that both the tree-level form factor \eqref{eq:tree-level-F3} and the four-gluon tree-level amplitude entering the one-loop cut are identical in any Yang-Mills theory. Theory-dependence will manifest itself at two and higher loops where the differences in matter content of the theories will become important. 


\section{Two-loop minimal form factors in $\N\!<\!4$ SYM}\label{Sec:4}

We now compute the minimal form factors $F_{\O_{\mathcal  S}}(1^+,2^+,3^+;q)$ and $F_{\O_{\mathcal  C}}(1^+,2^+,3^+;q)$ at two loops and in theories with less-than-maximal supersymmetry. 

\subsection{An  effective supersymmetric decomposition}
\label{susydecc}
There are two modifications one needs to take into account when decreasing the number of supersymmetries, $\N$, from the maximal value of $\N\!=\!4$. 
\begin{figure}[h]
	\centering
	\includegraphics[width=0.6\linewidth]{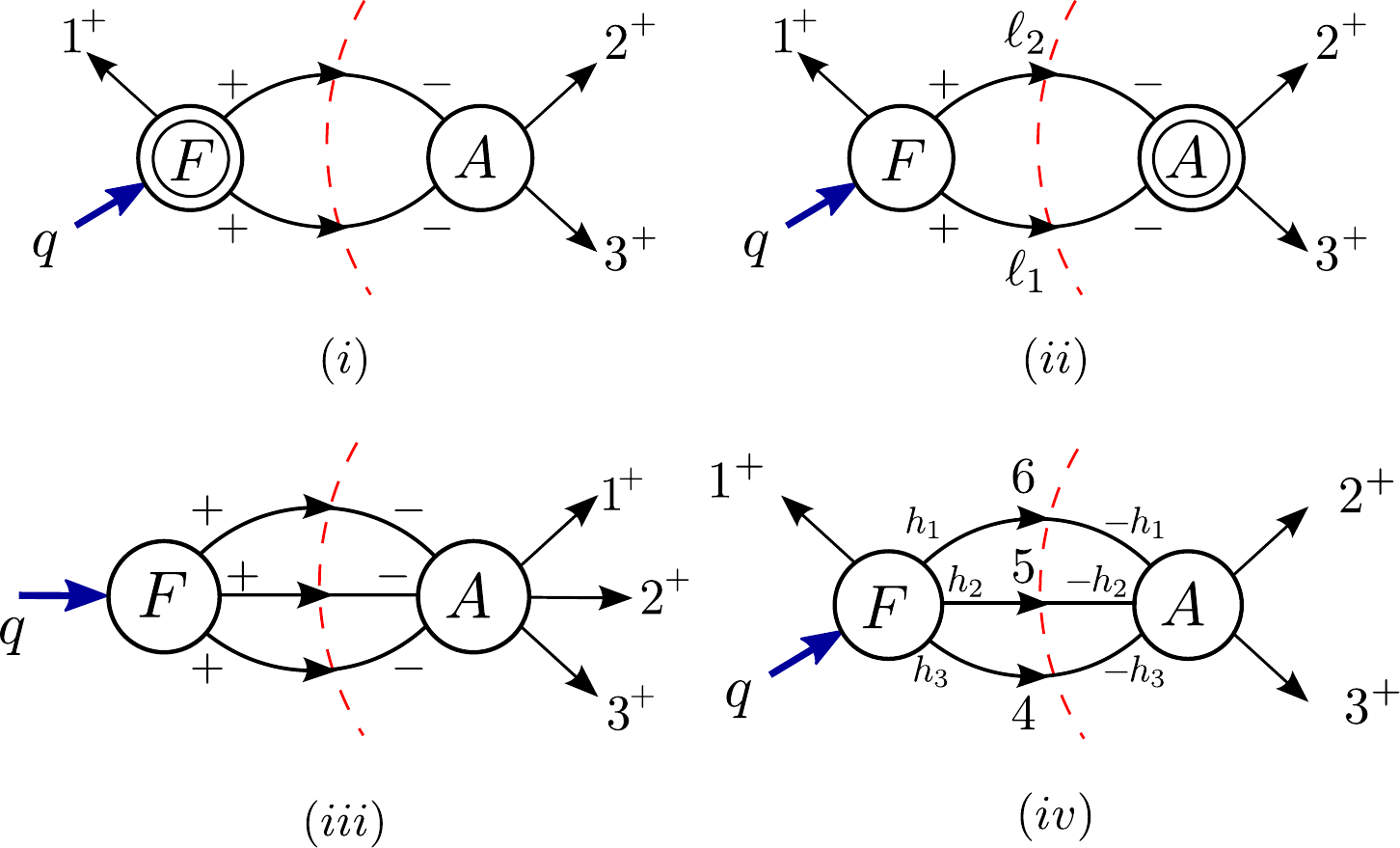}
\caption{\it Four unitarity cuts used to construct the integrand for the two-loop form factor of $F_{\O_\cS,\O_\cC}(1^+,2^+,3^+;q)$. Cut (i) and (iii) are both operator- and theory-independent. Cut (ii) is operator-independent, but theory dependent due to the presence of a one-loop sub-amplitude. Finally, cut (iv) probes both the specific operator and the theory, see also Table \ref{tab:cut-dependence}.}
\label{fig:allcuts}
\end{figure}

First, in computing the two-loop remainder functions the subtraction of the universal infrared divergences for theories with less-than-maximal supersymmetry must be substituted by a more general formula introduced by Catani \cite{Catani:1998bh}, featuring the non-zero beta function of the theory. 

Second, the two-loop integrand constructed in \cite{Part1} using the generalised unitarity cuts presented in Figure~\ref{fig:allcuts} above may receive contributions from  different states depending on the field content of the theory. The various  supersymmetric and non-supersymmetric theories  differ by the number of scalars and fermions in the vector multiplet. Hence, the key to understanding the difference between two-loop form factors in these  theories lies  in computing the individual contributions of scalars and fermions to the two- and three-particle   cuts shown in Figure~\ref{fig:allcuts}. 

\begin{figure}[h]
	\centering
	\includegraphics[width=0.30\linewidth]{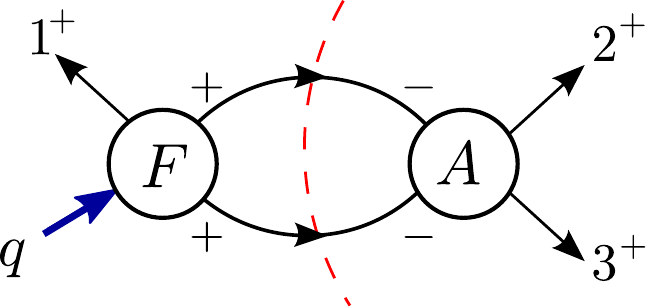}
	\caption{Two-particle cut of the one-loop form factor $F_{\O_{\mathcal  S},\OC}^{(1)}(1^{+},2^{+},3^{+};q)$.}
	\label{fig:1loopFF}
\end{figure}

However, inspecting the cuts in Figure \ref{fig:allcuts} carefully, it is clear that only $(ii)$ and $(iv)$ are sensitive to the field content of the theory since they feature a non-minimal form factor or a one-loop amplitude. Indeed, cut $(iii)$ involves only a tree-level form factor and an amplitude with gluons as external states, rendering it independent of the field content of the theory. 
Cut $(i)$ is slightly more subtle as it features a one-loop form factor which can in principle involve fermions and scalars running in the loop. For this particular configuration of external states, however, the cut of the one-loop form factor consists solely of tree-level quantities with gluons as external states, as shown in Figure~\ref{fig:1loopFF}.

 Thus we conclude that only cuts $(ii)$ and $(iv)$ are sensitive to the amount of supersymmetry. Even so, cut $(ii)$ depends on the field content only  through the one-loop amplitude, whose cut-constructible part receives additional contributions proportional to bubble integrals compared to the $\cN\!=\!4$ SYM case \cite{Bern:1994cg}. We will show this explicitly for different values of $\N$ in Section~\ref{sec:1loop-amp}. 
 
Finally, the last cut, $(iv)$, also depends on  the particular matter content due to   the  nontrivial sum over internal  fermions and scalars running in the loops. However, the story for the two operators $\cO_\cS$ and $\cO_\cC$ unfolds in two different ways. 
For  $\cO_\cC$, 
 the only possible matter-dependent contributions to cut $(iv)$  involve an internal state with a positive-helicity gluon and two adjacent scalars or fermions. Hence, the situation is entirely parallel to that of cut $(ii)$, in that the matter content dependence is restricted to one-loop sub-diagrams. This allows us,  for the case of $\cO_\cC$, to use a supersymmetric decomposition of the calculation as done in \cite{Bern:1994cg} for one-loop amplitudes. This is a remarkable and important simplification which does not apply to a generic two-loop amplitude computation. 
In the following we will  obtain the result of this cut as a function of $c_B$ (the number of complex scalar fields) and $c_F$ (the number of Weyl fermions) in each theory. This computation will be presented in detail in Section~\ref{sec:3-particle-cut-N-less-than-4}. 

The situation for $\cO_\cS$ is different because this operator contains additional terms giving rise to modifications to   tree-level form factors due to  the length-four terms inside $\cO_\cS$;  furthermore, $\cO_\cS$ depends on whether we consider  the $\cN\!=\!4$ or $\cN\!=\!2$  theory due to the state sum reduction. We also recall  that there is no distinction between the $\cO_\cC$ and  $\cO_\cS$ cases in $\cN\!=\!1$ SYM -- the only possible differences between the two operators are four-gluino terms, which cannot contribute to the process under consideration.

We briefly summarise in Table~\ref{tab:cut-dependence} what we know about the contributions from the individual cuts so far, and next we discuss modifications arising from the two- and three-particle cuts in turn.

\bgroup
\def\arraystretch{1.5}
\begin{table}[!h]
\centering
\begin{tabular}{cc|c|c|}
\cline{3-4}& &\bf Theory-independent? &\bf $\O_{\mathcal  S}$ same as $\O_{\mathcal  C}$?\\ \hline
\multicolumn{1}{|c}{\multirow{2}{*}{\bf Two-particle cut}}&\multicolumn{1}{ |c| }{\color{black} $(i):\,F^{(1)}\times A^{(0)}$}&\cmark&\cmark\\ \cline{2-4}
\multicolumn{1}{|c}{} & \multicolumn{1}{ |c| }{\color{black} $(ii):\,F^{(0)}\times A^{(1)}$} & \xmark& \cmark\\ \hline
\multicolumn{1}{|c}{\multirow{2}{*}{\bf Three-particle cut }} &\multicolumn{1}{ |c| }{\color{black} $(iii):\,q^2$-channel}&\cmark&\cmark\\ \cline{2-4}
\multicolumn{1}{|c}{ } & \multicolumn{1}{ |c| }{\color{black} $(iv):\,s_{23}$-channel} & \xmark& \xmark\\ \hline
\end{tabular}
\caption{\it Summary of the theory- and operator-dependence of the unitarity cuts of the two-loop form factor.}
\label{tab:cut-dependence}
\end{table}
\egroup

 \subsection{Modifications to the two-particle cut}
\label{sec:1loop-amp}
The two-particle cut with $F^{(0)}\times A^{(1)}$, presented in Figure \ref{fig:allcuts}$(ii)$ contains a four-point one-loop amplitude. If the matter content is changed compared to that of $\N\!=\!4$ SYM the amplitude will be modified by additional bubble integrals \cite{Bern:1994cg, Bern:1993mq, Bedford:2004nh}. Fortunately, for the four-point amplitude the modification is very simple. Explicitly, we have \cite{Bern:1994cg, ArkaniHamed:2008gz}
\begin{align}
A^{(1)}_{\N \leq 4}(\ell_1^-,\ell_2^-,2^+,3^+)\,=\,A^{(1)}_{\N=4}(\ell_1^-,\ell_2^-,2^+,3^+)-\beta_0\,A^{(1)}_{\N=1\,{\rm chiral}}(\ell_1^-,\ell_2^-,2^+,3^+)\,,
\end{align}
where $\beta_0$ is the first coefficient of the beta function of the theory in question (see Table \ref{tab:beta0} for its values in our conventions),  and 
\begin{align}
A^{(1)}_{\N=1\,{\rm chiral}}(\ell_1^-,\ell_2^-,2^+,3^+)\,=\, A^{(0)}(\ell_1^-,\ell_2^-,2^+,3^+)\times\pic{0.4}{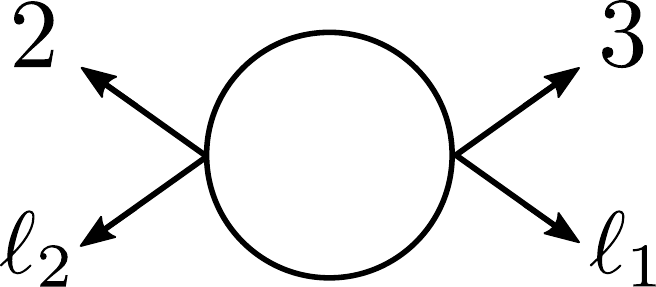}\,.
\end{align}
Once multiplied by the usual tree-level form factor \eqref{eq:tree-level-F3}, this additional contribution  gives rise to a new topology, absent in $\N\!=\!4$ SYM:
\begin{align}
\label{eq:ExtraSOTB}
\beta_0\,\frac{\Tr_+(1\ell_2\ell_1132)}{s_{12}s_{13}}\times\pic{0.4}{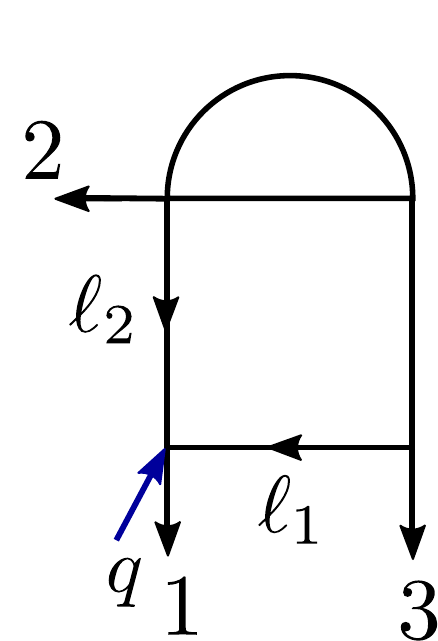}\,.
\end{align}
We note that this integral is free of any ambiguities as numerator terms involving powers of $\ell_1^2$ or $\ell_2^2$ would lead to scaleless integrals. Moreover, we do not expect to observe this integral in any of the other  cut channels we considered -- thus, 
we can simply add it to the integrand of the two-loop form factor. Finally, as indicated in Table \ref{tab:cut-dependence}, this cut is universal for both operators $\O_{\mathcal  S}$ and $\O_{\mathcal  C}$ and therefore its contribution to the integrands of both form factors is the same.

The important point we wish to make here is that, upon integral reduction, such an additional contribution can only produce two-loop integrals of sub-maximal transcendentality. As a consequence, the maximally transcendental part of the result remains unaltered by modifications of this cut, as already observed in \cite{Brandhuber:2017bkg}.

\subsection{Modifications to the three-particle cut}
\label{sec:3-particle-cut-N-less-than-4}

Having considered all modifications to two-particle cuts arising from studying different supersymmetric Yang-Mills theories, it remains  to inspect more closely  the individual contributions of scalars and fermions to the calculation of the $s_{23}$-channel three-particle cut, presented in Figure \ref{fig:allcuts}$(iv)$. We do this in detail for the component operator $\cO_\cC$, which is the  case compatible with a supersymmetric decomposition, as discussed earlier in Section \ref{susydecc}. 

Using the relevant expressions for tree-level form factors and amplitudes   explicitly quoted in (4.20)-(4.26) of  the companion paper \cite{Part1},  and leaving the $R$-symmetry multiplicities unspecified as $c_F$ for fermions and $c_B$ for scalars, after some manipulation we can bring all the scalar and fermion terms to a compact form:
\begin{align}
\begin{split}
\frac{\la46\ra}{\la 23\ra \la34\ra \la 62\ra}&\left[\frac{1}{s_{56}}\Big([1|54|1](-c_Fs_{45}+\frac{1}{2}\,c_Bs_{46})+[1|64|1](-c_Fs_{46}+\frac{1}{2}\, c_Bs_{45})\Big)\right.\\[5pt]
&\left.+\frac{1}{s_{45}}\Big([1|65|1](-c_Fs_{56}+\frac{1}{2}\,c_Bs_{46})+[1|64|1](-c_Fs_{46}+\frac{1}{2}\,c_Bs_{56})\Big)\right]\,.
\end{split}
\end{align}
We can then draw the corresponding integrals in this expression term-by-term:
\begin{align}\label{eq:SOTB1}
\mbox{\bf First term}&=\pic{0.4}{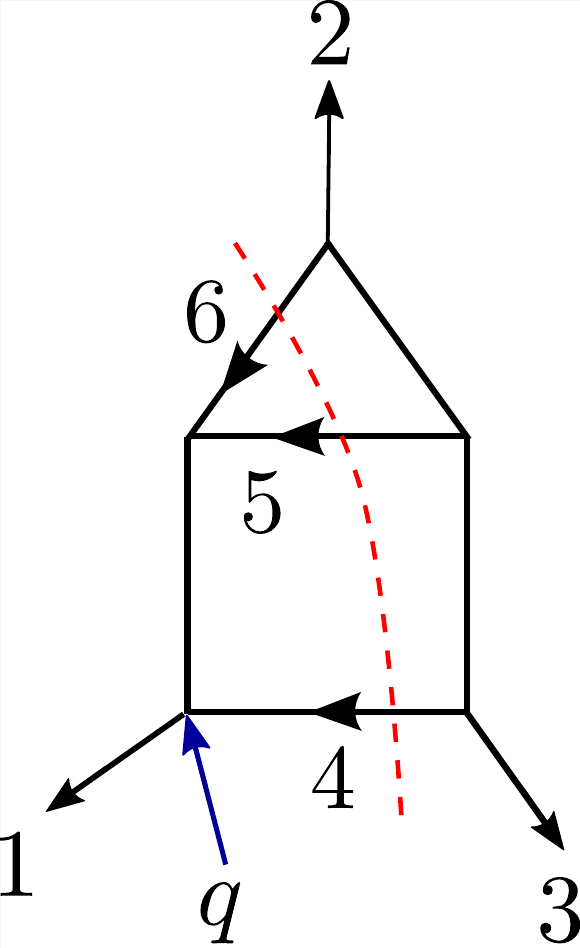}\times \frac{F^{(0)}_{\O_{\mathcal  S}}}{s_{12}s_{23}s_{31}}(-c_Fs_{45}+\frac{1}{2}\,c_Bs_{46})\,\Tr_+(26431541)\,,\\
\mbox{\bf Second term}&=\pic{0.4}{SOTLB-cut1}\times \frac{F^{(0)}_{\O_{\mathcal  S}}}{s_{12}s_{23}s_{31}}(c_Fs_{46}-\frac{1}{2}\,c_Bs_{45})\,\Tr_+(16413462)\,,\\
\mbox{\bf Third term}&=\pic{0.4}{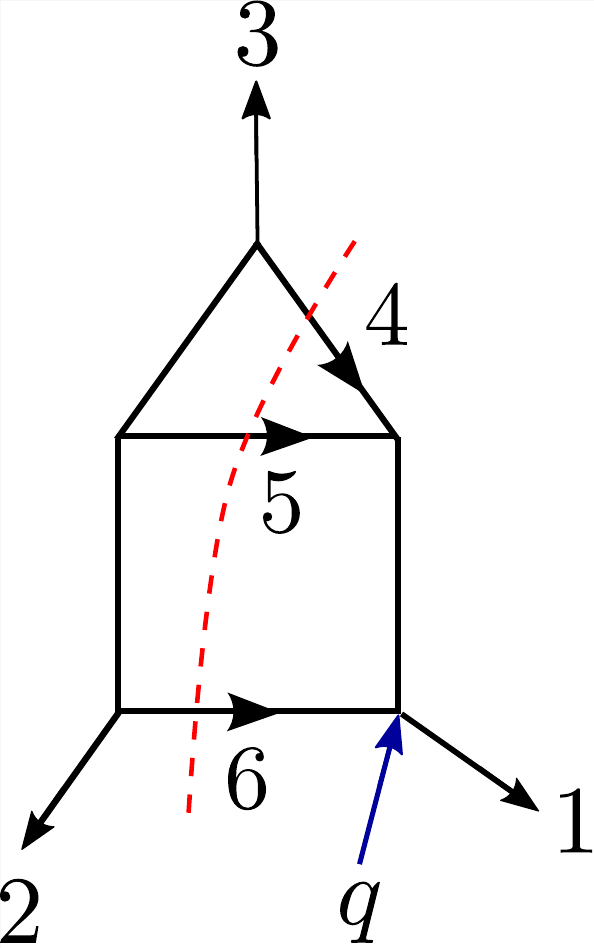}\times \frac{F^{(0)}_{\O_{\mathcal  S}}}{s_{12}s_{23}s_{31}}(c_Fs_{56}-\frac{1}{2}\,c_Bs_{46})\,\Tr_+(24631651)\,,\\\label{eq:SOTB4}
\mbox{\bf Fourth term}&=\pic{0.4}{SOTRB-cut1}\times \frac{F^{(0)}_{\O_{\mathcal  S}}}{s_{12}s_{23}s_{31}}(-c_Fs_{46}+\frac{1}{2}\,c_Bs_{56})\,\Tr_+(14613462)\,.
\end{align}
The reduction of these integrals with complicated-looking numerators leads to surprisingly simple results. For example, the term in  \eqref{eq:SOTB1} reduces to
\begin{align}
\hspace{-15pt}\begin{split}\label{eq:reductionfermscal}
&-\frac{c_B(6d+ 4d^2- 5d^3 + d^4) + c_F(40d  - 40 d^2 + 14 d^3 -  2d^4)}{24 (d-4)^2 (d-3) (d-2) ( d-1) (p_2\cdot p_3)}\times\!\pic{0.65}{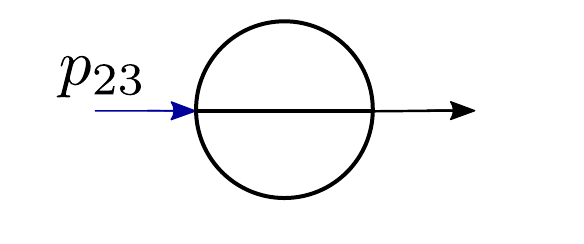}\\
&-\frac{c_B(-96+ 137d- 53d^2+  6d^3) +c_F(- 96  + 84 d  - 12d^2 )}{12 (d-4) (d-1) ( 3 d-8)}\times\!\pic{0.6}{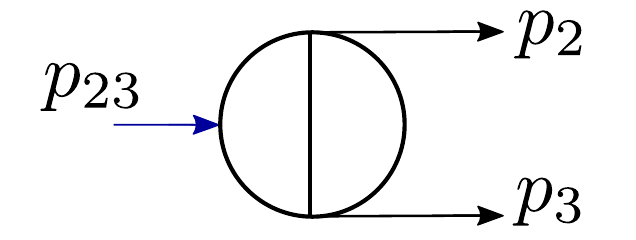}\ ,
\end{split}
\end{align}
which, after explicit evaluation, turns out to be of transcendentality three and lower. Again, we see that regardless of the number of fermions and scalars present in the theory, their contribution is submaximal in transcendentality. As a result, we arrive at the important conclusion that {\it the maximally transcendental part of the two-loop form factor is universal for Yang-Mills theories with any amount of supersymmetry, as anticipated in 
\cite{Brandhuber:2017bkg}.
As far as QCD is concerned the same conclusion holds -- the presence of fermions in the fundamental representation only alters group theory factor and does not lead to  new types of integrals. }

A final  observation is in order. In \eqref{eq:reductionfermscal}, which is the result of the integral reduction of \eqref{eq:SOTB1}, we see two two-loop master topologies arising. While the first topology is perfectly consistent with the cut we are considering -- three-particle in the $s_{23}$-channel, the second topology arising from the reduction does not have a cut in this channel. 
Demanding consistency of the cut and the topology it gives rise to, we conclude that such contribution is inconsistent and therefore  we drop it from the result.

\section{Remainder functions in $\N\!<\!4$ SYM}\label{Sec:7}
\subsection{Catani form factor remainder and renormalisation}
For theories with $\N<4$ supersymmetry, which have non-vanishing beta function, one must take into account  renormalisation. 
Catani's remainder is expressed in terms of renormalised quantities, and hence we need to first discuss how these are related to the bare quantities (which is what we calculate). 

We begin by noting  that in the $\overline{\text{MS}}$ scheme, the bare coupling constant as a function of the renormalised coupling at a scale $\mu$, denoted by $a(\mu)$, is given by \cite{Catani:1998bh}
\begin{align}
\label{eq:alpharen}
a^{U}S_\eps=\left(\frac{\mu}{\mu_0}\right)^{2\eps}a(\mu)\left[1-a(\mu)\frac{\beta_0}{\eps}+a^2(\mu)\left(\frac{\beta_0^2}{\eps^2}-\frac{\beta_1}{2\eps}\right)\right]+\O(a^4(\mu))\ , 
\end{align}
where  $S_\eps\coloneqq(4\pi)^\eps e^{-\gamma_E \eps}$ and  $\beta_0$, $\beta_1$ are the first two coefficients of the beta function for the 't Hooft coupling,
\begin{align}
\beta(a(\mu))\,\coloneqq \,\mu\frac{\partial a(\mu)}{\partial \mu}\, , 
\end{align}
and $\beta(a) = - 2 a \eps -2 a^2 \beta_0 - 2 a^3 \beta_1 + \cO(a^4)$.
Note that we define the  't Hooft coupling as
\begin{align}\nonumber
a \,=\, {g^2 N \over  (4 \pi)^2}\,  .
 \end{align}
 The values of  $\beta_0$ are well-known for any $SU(N)$ gauge theory \cite{Gross:1973ju}\begin{align}
\beta_0\,=\,
 \frac{11}{3}-\frac{1}{6}\sum_i {C_i\over N} -\frac{2}{3}\sum_j {\tilde C_j\over N}\,,
\end{align}
where the first sum is over all real scalars and the second sum over all Weyl fermions with quadratic Casimirs $C_i$ and $\tilde C_j$ respectively. Since we are dealing with Yang-Mills theories without matter, all fields are in the adjoint representation and thus $C_j\!=\!\tilde{C}_j\!=\!N$. For convenience, we  list in Table~\ref{tab:beta0} below the values of $\beta_0$ and $\beta_1$  for $\N\!=\!0,1,2,4$.
\begin{table}[h]
	\centering
	\begin{tabular}{|c|c|c|c|c|}
		\hline 
${\N}$ & $\#$ \bf real scalars & $\#$ \bf Weyl fermions  & $ \beta_0$& $ \beta_1$\\
		\hline 4 & 6 & 4 & 0& 0 \\
		\hline 2 & 2 & 2 & 2& 0 \\
		\hline 1 & 0 & 1 & 3& 6 \\
		\hline 0 & 0 & 0 & 11/3 & 34/3\\ \hline
	\end{tabular}
	\caption{Field content and values for $\beta_0$ and $\beta_1$ for Yang-Mills theories with $\N=0,1,2,4$ supersymmetry.}
	\label{tab:beta0}
\end{table}

Form factors can be interpreted as amplitudes in theories where an additional operator $\O$ with coupling $\lambda$ has been added to the Lagrangian. If the operator is multiplicatively
renormalisable,  then the coupling $\lambda$ is renormalised as%
\footnote{We will find later in \eqref{eq:consistency} that the quantity $
 \textcolor{black}{\rho_2}
$ appearing in \eqref{eq:lambdaren} can be re-expressed in terms of   $\gamma_0$ and  $\beta_0$ as a simple consequence   of $\mu \, \partial\lambda^U/ \partial \mu = 0$.}
\begin{align}
\label{eq:lambdaren}
\lambda^{U}=\lambda(\mu)\left[1-a(\mu)\frac{\gamma_0}{\eps}+\frac{a^2(\mu)}{2}\left(\frac{
 \textcolor{black}{\rho_2}
}{\eps^2}-\frac{\rho_1}{\eps}\right)\right]+\O(a^4(\mu))\ .
\end{align}
Thus, we can write a renormalised form factor in two ways, either as functions of bare or renormalised quantities. Up to two loops we have
\begin{align}
\begin{split}
F^{R}_\O &= \lambda(\mu) \left[(F^{R}_\O)^{(0)} + a(\mu)(F^{R}_\O)^{(1)}+a^2(\mu)(F^{R}_\O)^{(2)}\right]+\O\big(a^4(\mu)\big)\\[8pt]
&=\lambda^U \left[(F^{U}_\O)^{(0)} + a^U(F^{U}_\O)^{(1)}+(a^U)^2(F^{U}_\O)^{(2)}\right]+\O\big((a^U)^4\big)\ .
\end{split}
\end{align}
Using \eqref{eq:alpharen} and \eqref{eq:lambdaren} in the above equation, we can solve for the renormalised form factors in terms of the bare ones, arriving at the following relations: 
\begin{align}
\label{eq:FFren0}
(F^{R}_\O)^{(0)} &= (F^{U}_\O)^{(0)} \ ,\\\label{eq:FFren1}
(F^{R}_\O)^{(1)} &= \left(\frac{\mu}{\mu_0}\right)^{2\eps}\frac{(F^{U}_\O)^{(1)}}{S_\eps}- \frac{\gamma_0}{\eps}(F^{U}_\O)^{(0)} \ ,\\\label{eq:FFren2}
(F^{R}_\O)^{(2)} &= \left(\frac{\mu}{\mu_0}\right)^{4\eps}\frac{(F^{U}_\O)^{(2)}}{S_\eps^2} -\frac{1}{\eps}\left[(\beta_0+\gamma_0)\left(\frac{\mu}{\mu_0}\right)^{2\eps}\frac{(F^{U}_\O)^{(1)}}{S_\eps}+\frac{\rho_1}{2} (F^{U}_\O)^{(0)}\right] +(F^{U}_\O)^{(0)}\frac{
 \textcolor{black}{\rho_2}
}{2\eps^2}	\ , 
\end{align}
where the superscripts  $U$ and $R$ stand for unrenormalised and renormalised. 

An important comment on operator mixing is in order here. As fully discussed in Section~5.1 in the companion paper \cite{Part1} for the case of $\mathcal{N}\!=\!4$  SYM the operator $\Tr (F^3)$ and its supersymmetric completion mix with the operator $\O_{\mathcal M} \sim \Tr(D_\mu F_{\nu\rho} D^\mu F^{\nu\rho})$. The mixing manifests itself in the non-vanishing off-diagonal term of the mixing matrix in Eqn.~(5.12) in \cite{Part1}. However this term is directly related to the UV divergence of the sub-minimal two-loop form factor of $\Tr (F^3)$ computed in Section~4.7 of \cite{Part1}. Importantly, this quantity turns out to be  theory independent since it only gets a contribution from a triple cut involving a minimal three-point form factor and a five-particle gluon amplitude (see Figure~16 of \cite{Part1}). Hence, for all practical purposes this mixing effect is identical in all cases  and, hence, the corresponding UV divergence can be removed universally. The remaining UV divergences can then be removed by multiplicative renormalisation as described above.

We are now ready to use  these expressions  and define finite remainders. Having removed ultraviolet divergences, the final step is to remove the universal infrared ones. 
At  one loop, the finite remainder  is defined  as 
\beq
\label{Catani1loop}
\cR^{(1)}(\epsilon) \coloneqq 
(\mathcal{F}^{R}_\O)^{(1)} \, - \,  \, I^{(1)} (\eps)
\, , 
\eeq
where $\mathcal{F}^{R\, (L)}_{\cO}\coloneqq (F_\O^R)^{(L)} / (F_\O)^{(0 )}$,  
$(F^{R}_\O)^{(1)}$ is the one-loop renormalised remainder defined in \eqref{eq:FFren1}, and the expression for $I^{(1)} (\eps)$  for $n$ gluons is 
\cite{Giele:1991vf,Kunszt:1994np,Catani:1996jh, Catani:1996vz}
\beq
I^{(1)} (\eps)\ = \ - {e^{\eps \gamma} \over \Gamma (1 - \eps)} \Big( {1\over \eps^2} + {\beta_0\over 2 \eps}\Big)\sum_{i=1}^n 
 \left( - {s_{i i+1} \over \mu^2} \right)^{- \eps} 
\ . 
\eeq
Next we  introduce  the two-loop  Catani remainder \cite{Catani:1998bh}  
in the the formulation of \cite{Bern:2004cz}. This is given by 
\begin{align}\nonumber\label{eq:Catani}
\mathcal{R}^{(2)}(\epsilon)\,\coloneqq\,({\mathcal{F}}^R_\O)^{(2)}(\epsilon)&-\hf \left[(\mathcal{F}^{R}_\O)^{(1)}(\epsilon)\right]^2
+\frac{\beta_0}{\epsilon}(\mathcal{F}^{R}_\O)^{(1)}(\epsilon)\\
&-e^{-\gamma_E\epsilon}\,\frac{\Gamma(1-2\epsilon)}{\Gamma(1-\epsilon)}(\mathcal{F}^{R}_\O)^{(1)}(2\epsilon)\left(\frac{\beta_0}{\epsilon}+K\right)+\frac{n\,e^{\gamma_E\epsilon}}{4\epsilon\,\Gamma(1-\epsilon)}\,H^{(2)}
\, , 
\end{align}
where $n$ is the number of legs ($n=3$ for the case in question). 
The particular values of  $K$  and $H^{(2)}$ required in order to guarantee the infrared finiteness of the remainder are
\begin{align}\label{eq:K}
K_{\rm SYM}\,&=\,2\left[(4-\N)-\zeta_2\right]\,,
\\\label{eq:H}
H^{(2)}_{\rm SYM}\,&=\,2\,\zeta_3+\frac{(4-\N)}{2}\,\zeta_2\,, 
\end{align}
where  $\cN>0$ is the number of supersymmetries.%
\footnote{This choice is not unique however. Compared with the conventions of   (A.27) and (A.32) of \cite{Bern:2004cz} for $\cN\!=\!1$ SYM,  we have shifted an  $\cO( \eps)$ term from  $K_{\rm SYM}$ to  $H^{(2)}_{\rm SYM}$. Therefore the latter   is  shifted  by a rational constant with respect to \cite{Bern:2004cz}.}

Away from $\N\!=\!4$ SYM,  the values of parameters $\gamma_0$, $\rho_1$ and $\rho_2$ appearing in \eqref{eq:FFren1} and \eqref{eq:FFren2} are not yet determined. 
We are now going to fix $\gamma_0$, which in turn is related to the one-loop anomalous dimensions of the operators. We will fix the remaining parameters in the next sections as they require two-loop data. 

The constant $\gamma_0$ can be determined by requiring the finiteness of the one-loop remainder \eqref{Catani1loop} with the one-loop unrenormalised minimal form factor \eqref{eq:one-loop-result} as an input. This leads to the relation
\beq
\gamma_0 \ = \ - 6 \, + \, {3\over 2} \, \beta_0
\ . 
\eeq 
Note that this result is the same for the two operators $\cO_{\cS}$ and  $\cO_{\cC}$. 
The one-loop anomalous dimension of the corresponding operators $\gamma^{(1)}_{\cO_{\cS,\cC}}$ is simply 
\beq
\gamma^{(1)}_{\cO_{\cS,\cC}} \ = \ - 2 \, \gamma_0 \ = \ 12 \, -\,  3 \, \beta_0
\ . 
\eeq
In pure Yang-Mills $\beta_0 = 11/3$ and we get $\gamma^{(1)}_{\cO_{\cS}} = 1$, in agreement with \cite{Ferretti:2004ba}. For $\cN\!=\!4$ we get $\gamma^{(1)}_{\cO_{\cS}} = 12$, which is also the correct result \cite{Anselmi:1996mq,Bianchi:1999ge}.

\subsection{$\N\!=\!2$ SYM}

In this section we  evaluate the two-loop form factors and the Catani remainder functions of the operators $\O_{\mathcal  S}$ and $\O_{\mathcal  C}$ in $\N\!=\!2$ SYM.

\subsubsection{The $\N\!=\!2$ SYM form factors}

As indicated by the summary in Table~\ref{tab:cut-dependence}, we  need to reconsider two types of cuts as they are theory-dependent: the two particle cut involving a one-loop amplitude and the three-particle cut in the $s_{23}$-channel. 

 There are two possible ways of finding the contribution of the $s_{23}$-channel three-particle cut to the two-loop integrand in $\N\!<\!4$ SYM. We can either follow the strategy described in Section 4 of \cite{Part1} and solve this cut numerically, or we can use the result for $\N\!=\!4$ SYM and appropriately subtract the contributions of scalars and fermions described in Section \ref{sec:3-particle-cut-N-less-than-4}. In the case of $\N\!=\!2$ SYM we subtract the contribution of 2 Weyl fermions and 4 real scalars from the $\N\!=\!4$ SYM integrand, which amounts to subtracting the integral topologies \eqref{eq:SOTB1}-\eqref{eq:SOTB4} with $c_F\!=\!2$ and $c_B\!=\!4$.  We have performed the calculation using  both methods, arriving at the same result. For convenience, we present below the outcome of the first method. 

\begin{figure}[!h]
\centering
\includegraphics[width=0.3\textwidth]{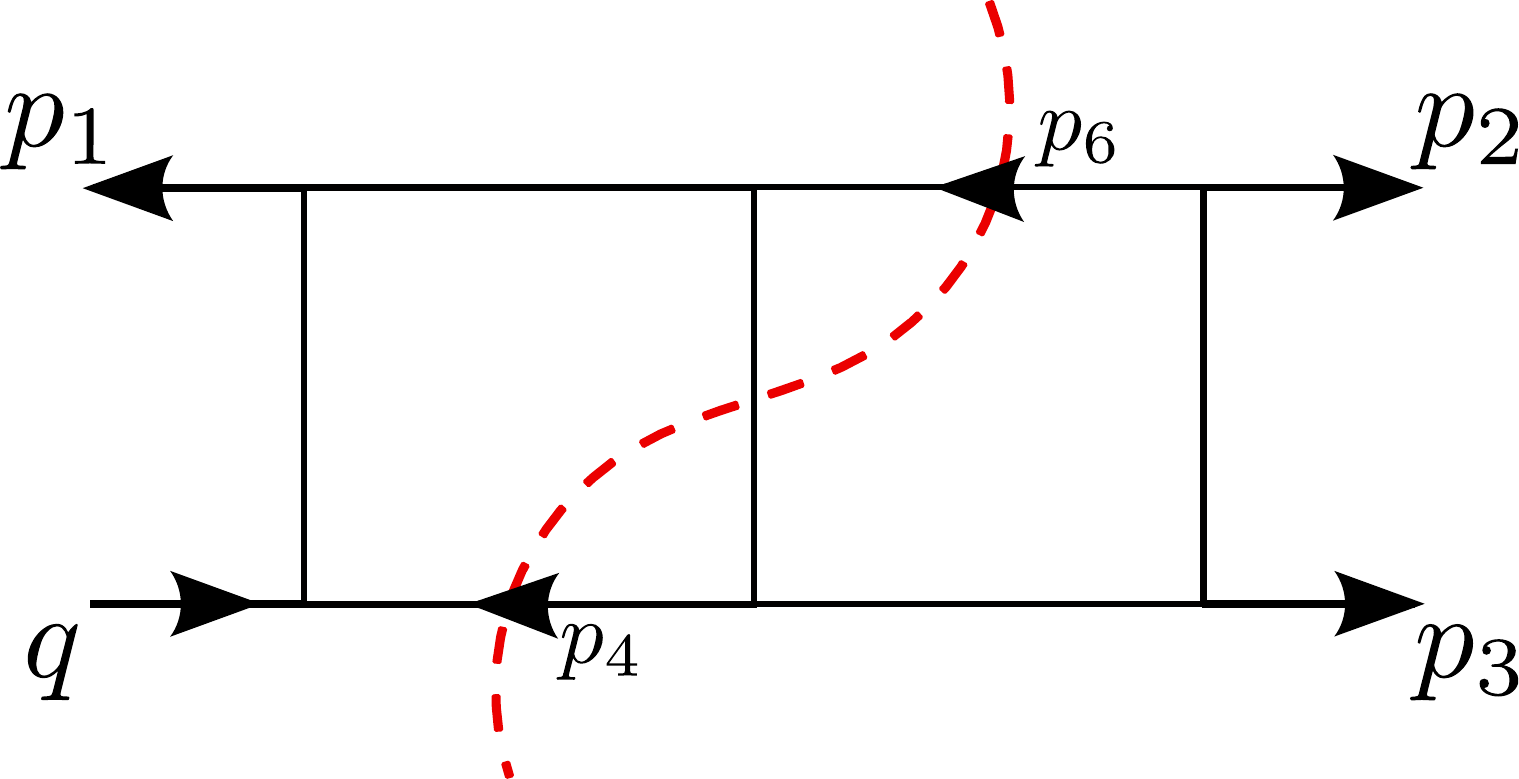}
\caption{\it One of the cuts of the maximal topology used to solve the $s_{23}$-channel triple cut. Note that $p_6$ is part of a one-loop sub-amplitude.}
\label{fig:power-counting}
\end{figure}

The procedure follows that of Section 4 of \cite{Part1}, with an important  modification of  the power counting  imposed on the numerator loop momenta. Specifically, the no-triangle property  of  $\N\!=\!4$ SYM strongly restricts the power counting of the  loop momenta belonging to a one-loop sub-amplitude. For example, for the cut topology presented in Figure \ref{fig:power-counting}, $p_6$ cannot feature in the numerator since the sub-amplitude can only contain scalar boxes. 

In  $\N\!<\!4$ SYM  the no-triangle property does not apply, and  $p_6$ can now appear  in the numerator. Solving for the $\N\!=\!2$ SYM integrand, we indeed observe new integral topologies which were previously forbidden by the no-triangle property of $\N\!=\!4$~SYM. These are shown as $I_{13}$ and $I_{14}$ in Table \ref{tab:two-loop basis}. The last topology, $I_{15}$ arises from the one-loop amplitude with $\N<4$ supersymmetry, \emph{cf.} \eqref{eq:ExtraSOTB}.

\begin{table}[!h]
\centering
\begin{tabular}{cccc}
$\pic{0.4}{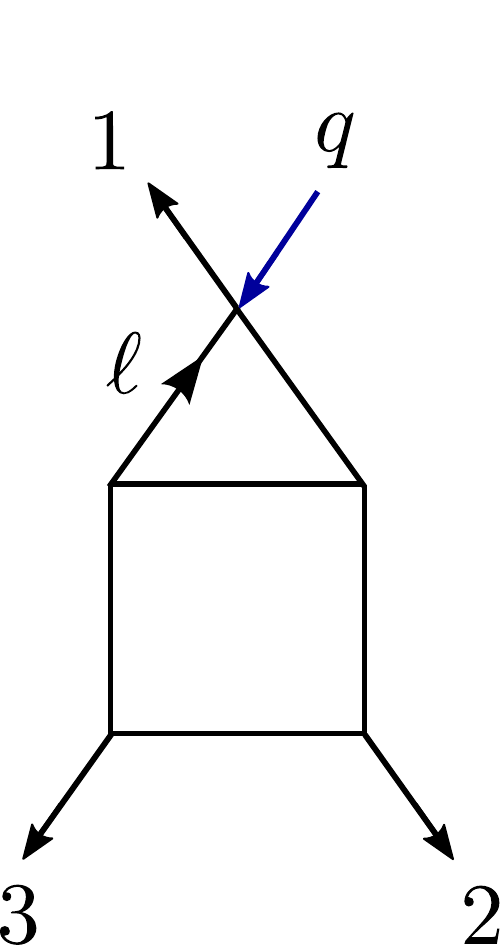}$ & $\pic{0.4}{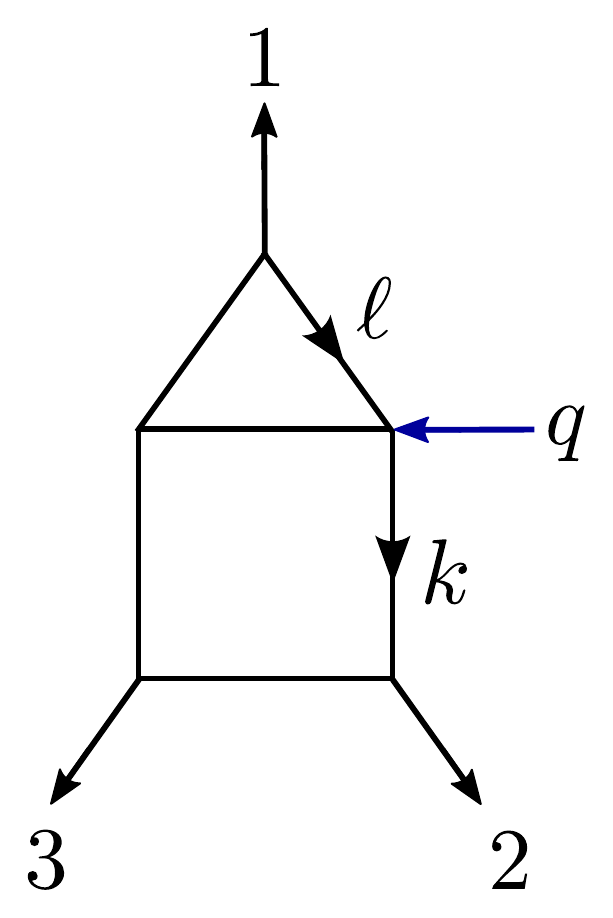}$ & $\pic{0.4}{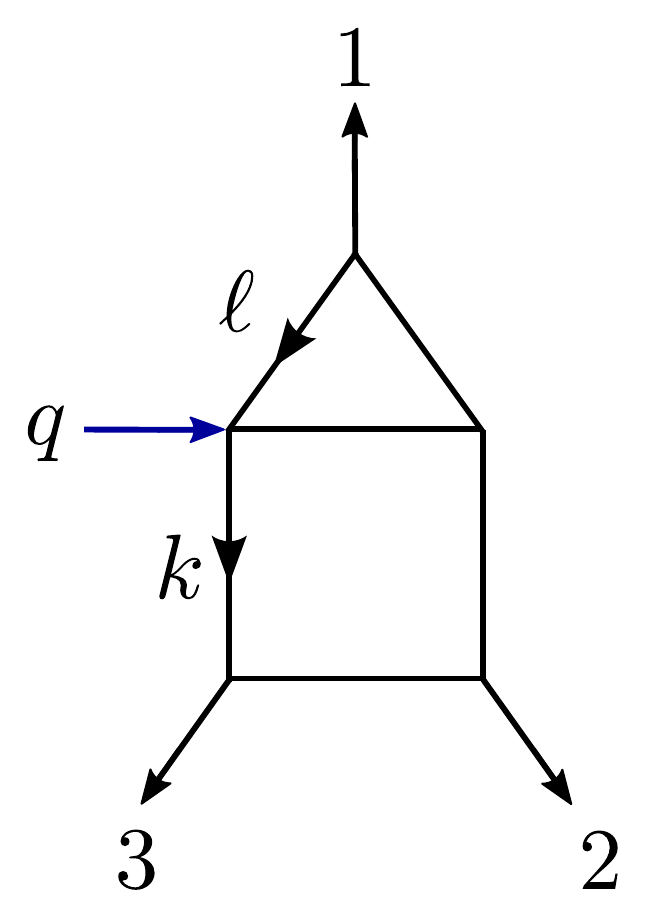}$ & $ \pic{0.4}{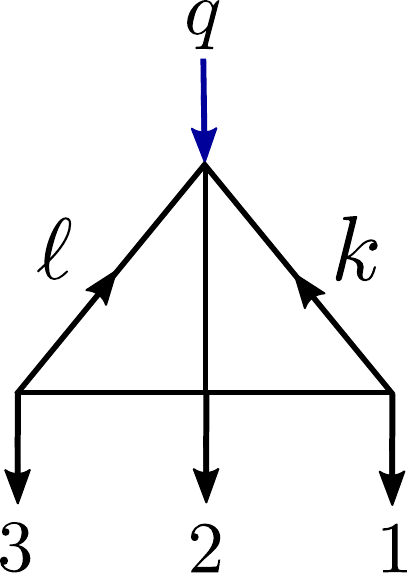} $ \\[30pt]
$I_1$ & $I_2$ &$I_3$ &$I_4$ \\[10pt]
$\pic{0.4}{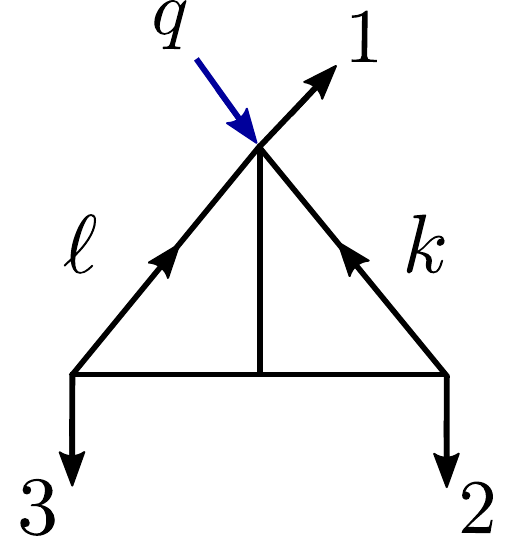} $ & $\pic{0.4}{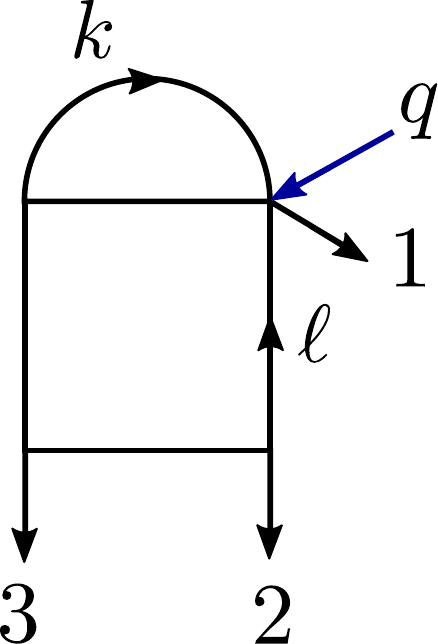}$ & $\pic{0.4}{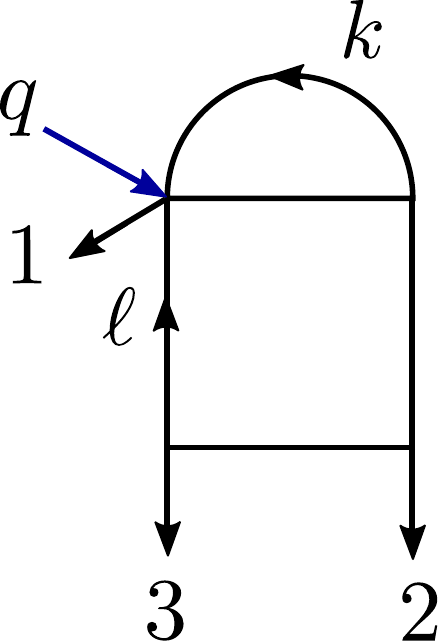}$ & $\pic{0.4}{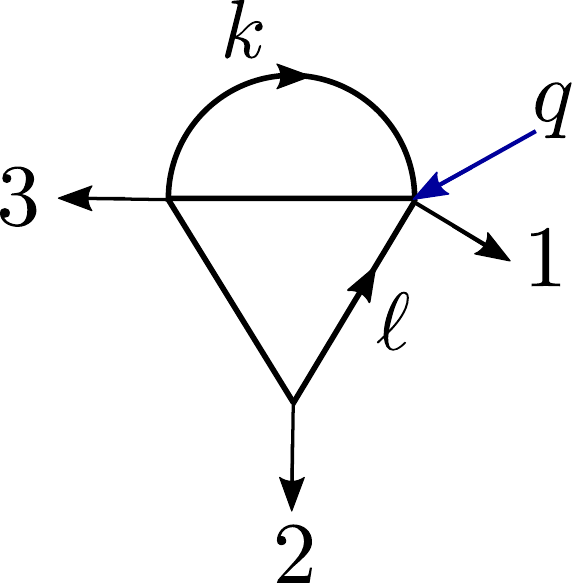}$\\[40pt]
$I_5$ & $I_6$ &$I_7$ &$I_8$ \\[10pt]
$\pic{0.4}{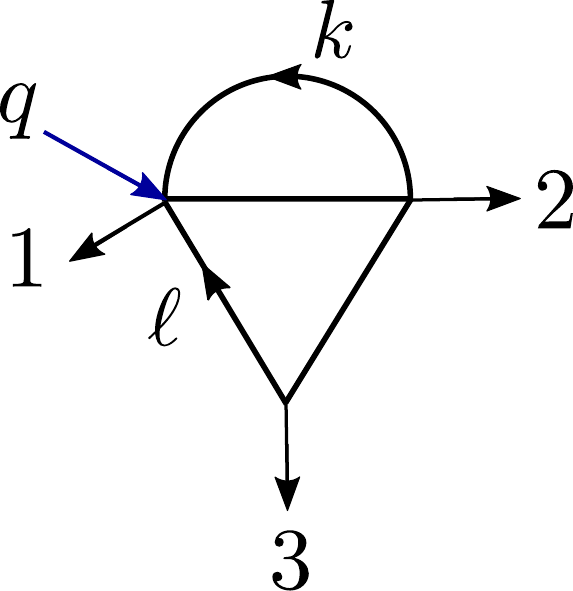} $ & $\pic{0.4}{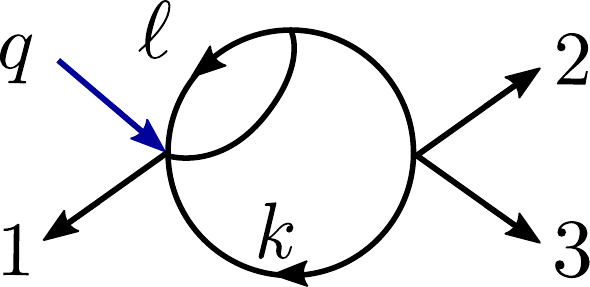} $ & $ \pic{0.4}{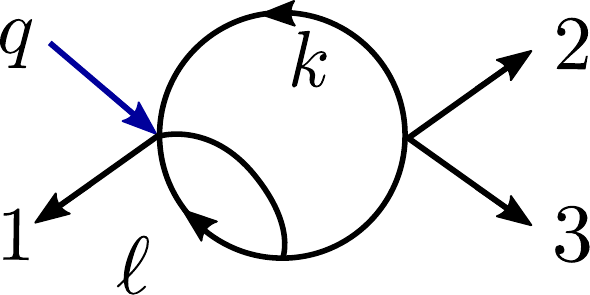} $ & $ \pic{0.4}{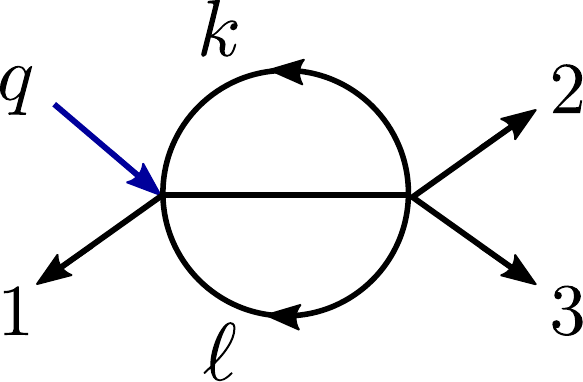}$\\[30pt]
$I_9$ & $I_{10}$ &$I_{11}$ &$I_{12}$ \\[10pt]
$\pic{0.4}{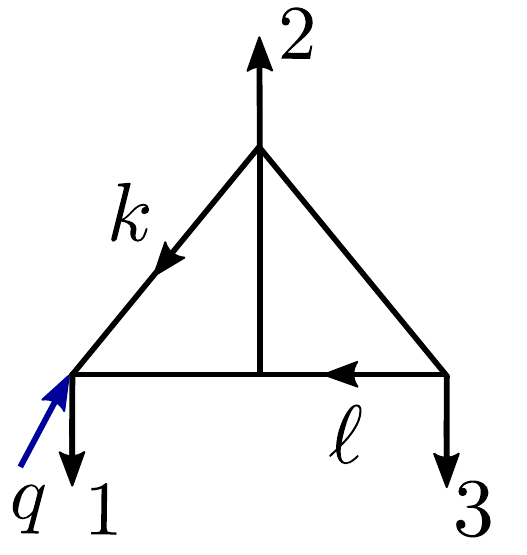} $ & $\pic{0.4}{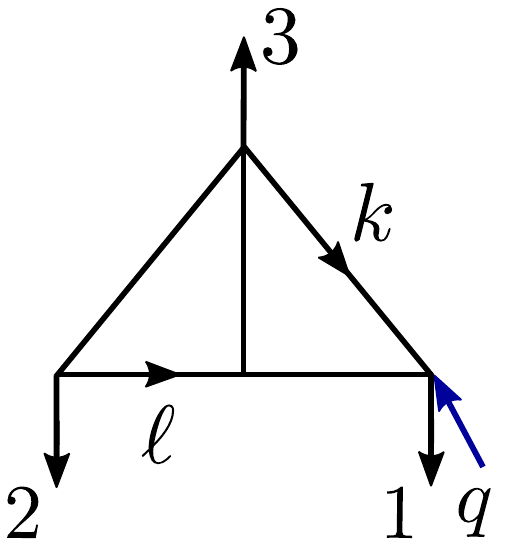} $ & $ \pic{0.4}{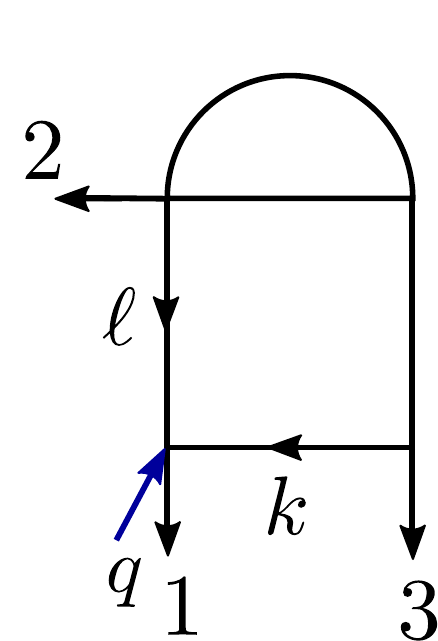} $ & \\[40pt]
$I_{13}$ & $I_{14}$ &$I_{15}$ & \\[10pt]
\end{tabular}
\caption{\it Integral basis for the two-loop form factor $F_{\O_{\mathcal  S},\OC}^{(2)}(1^{+},2^{+},3^{+};q)$ in $\N\!<\!4$ SYM.}
\label{tab:two-loop basis}
\end{table}

The full integrand for the two-loop form factor of $\O_{\mathcal  S}$ computed in $\N\!=\!2$ SYM, including the additional contributions from the modified two- and three-particle cuts, can be expressed in terms  the $\N\!=\!4$ SYM result plus an offset term: 
\begin{align}
F^{(2)}_{\N=2\,\O_{\mathcal  S}}\,=\,F^{(2)}_{\N=4\, \O_{\mathcal  S}}+\Delta_{\N=2\,\O_\mathcal{S}}\,,\qquad \Delta_{\N=2\,\O_\mathcal{S}}\,=\, \sum_{i=5}^{15} N'_i \times I_i\,,
\end{align}
with the numerators presented in \eqref{eq:ff-emeryN2N4} and the integrals  listed in Table \ref{tab:two-loop basis}. Note in particular the appearance of two new topologies in Table \ref{tab:two-loop basis}, and denoted as $I_{13}$ and $I_{14}$. As discussed in Section \ref{sec:1loop-amp}, the modification identified from  two-particle cuts is directly added to the integrand and is denoted as topology $I_{15}$. Similarly, the full integrand for the two-loop form factor of $\O_{\mathcal  C}$ computed in $\N\!=\!2$ SYM can be expressed as
\begin{align}\label{eq:resulf-differenceN2}
F^{(2)}_{\N=2\,\O_{\mathcal  C}}\,=\,F^{(2)}_{\N=4\, \O_{\mathcal  C}}+\Delta_{\N=2\,\O_{\mathcal  C}}\,,\qquad \Delta_{\N=2\,\O_{\mathcal  C}}\,=\, \sum_{i=5}^{15} \hat N_i \times I_i\,,
\end{align}
with the numerators presented in \eqref{eq:ff-cptN4N4}.

Having obtained the integrand for the two-loop form factors of $\O_\cS$ and $\O_\cC$ in $\N\!=\!2$ SYM, we follow the usual procedure of reduction to master integrals with the help  of {\tt LiteRed} \cite{Lee:2012cn, Lee:2013mka}  and evaluation using the known expressions of the master integrals of \cite{Gehrmann:1999as,Gehrmann:2000zt}. 

\subsubsection{The $\cN\!=\!2$ SYM remainders}
\label{Sec:N=2rem}

We now evaluate the two-loop remainder function given in \eqref{eq:Catani} for the operators 
$\O_\cS$ and $\O_\cC$, using the renormalised form factors \eqref{eq:FFren0}--\eqref{eq:FFren2} as  input. 

The first observation to make is that demanding the finiteness of the two-loop remainder, we can  fix the parameters appearing in the renormalised expressions, with the results:
\begin{align}\label{valuesN2}
\gamma_0\,=\,-3\,,\qquad \textcolor{black}{\rho_2}\,=\,3\,,\qquad \rho_{1\, , \O_{\mathcal  S}}\,=\,-2\,,\qquad \rho_{1\, , \O_{\mathcal  C}}\,=\,-3\,.
\end{align}
Next we move on to the finite $\cN\!=\!2$ SYM remainder. In order to present  it efficiently and at the same time highlight its main features, in Table~\ref{tab:difference-remainders} below we 
quote the difference between the $\cN\!=\!2$  and $\N\!=\!4$ SYM remainders, 
slice by slice in transcendentality degree.%
\footnote{For the reader's convenience we also write in Appendix \ref{Sec:N=4recap} the complete $\N\!=\!4$ SYM remainder.}
\bgroup
\def\arraystretch{1.5}
\begin{table}[h] 
	\centering
	\begin{tabular}{|c|c|c|}\cline{1-3}
	\hline
\multicolumn{1}{|c|}{\bf Transc.} & $\cR^{(2)}_{\N=2\,\O_{\mathcal  S}}-\cR^{(2)}_{\N=4\,\O_{\mathcal  S}}$& $\cR^{(2)}_{\N=2\,\O_{\mathcal  C}}-\cR^{(2)}_{\N=4\,\O_{\mathcal  C}}$ \\\hline
		\multicolumn{1}{|c|}{4}&0&0\\
		\multicolumn{1}{|c|}{3}&$-\frac{5}{2}\,\zeta_2[\log(uvw)+3\log(-q^2)]-\frac{11}{2}\zeta_3$&$-\frac{5}{2}\,\zeta_2[\log(uvw)+3\log(-q^2)]-\frac{11}{2}\zeta_3$\\
		\multicolumn{1}{|c|}{2}&$18\,\zeta_2$&$18\,\zeta_2$\\
		\multicolumn{1}{|c|}{1}&$\frac{14}{3}[\log(uvw)+3\log(-q^2)]$&$3[\log(uvw)+3\log(-q^2)]$\\
		\multicolumn{1}{|c|}{0}&$-\frac{65}{2}$&$-\frac{45}{4}$\\\hline
	\end{tabular}	
	\caption{\it Difference between two-loop Catani remainders of operators $\O_{\mathcal  S}$ and $\O_{\mathcal  C}$ when calculated in $\N\!=\!4$ and $\N\!=\!2$ SYM, split by transcendentality degree. }
	\label{tab:difference-remainders}
\end{table}
\egroup

Table \ref{tab:difference-remainders}  immediately shows the main feature of our result: it is almost identical to that of the remainder obtained in $\cN\!=\!4$ SYM! In more detail: 
\begin{itemize}
\item[{\bf 1.}]
The transcendentality-four slices of the remainders for $\O_{\mathcal  S}$ and $\O_{\mathcal  C}$ are identical and equal to that in the $\N\!=\!4$ SYM theory, {\it i.e.}~this quantity is universal, with the universality extending also to pure Yang-Mills and QCD \cite{Brandhuber:2017bkg}. 
\item[{\bf 2.}]
The difference between the remainders of operators when computed in $\N\!<\!4$ and $\N\!=\!4$ SYM is limited to a small number of terms as detailed in the table. Recalling the result of \cite{Brandhuber:2017bkg} for the $\N\!=\!4$ SYM remainder, also quoted in Appendix \ref{Sec:N=4recap}, we see that that expression contains ``pure" terms, {\it i.e.} purely transcendental functions, as well as ``non-pure" terms, which have rational prefactors. For instance, at transcendentality three we  found the prefactors  
\beq
\left\{{u\over v}, \, {v\over u}, \,{v\over w}, \, {w\over v}, \, {u\over w}, \, {w\over u}\right\}
\ , 
\eeq 
while at transcendentality two the list of prefactors is 
\beq
\left\{{u^2\over v^2}, \, {v^2\over u^2}, \, {u^2\over w^2}, \, {v^2\over w^2}, \, {w^2\over u^2}, \, {w^2\over v^2}\right\}\ .
\eeq
Strikingly, such non-pure terms in the $\cN\!=\!2$ SYM remainder are exactly the same as in $\cN\!=\!4$ SYM quoted in \eqref{eq:non-pure-3} and \eqref{eq:non-pure-2}.  As Table  \ref{tab:difference-remainders} shows, only pure logarithms, and $\zeta_2$ and $\zeta_3$ terms appear in the difference, without any rational prefactor. In \cite{Part1} it was shown that  these  rational  prefactors in the $\cN\!=\!4$ SYM result do not lead to unphysical soft/collinear singularities in the remainder function. That discussion applies also to the present context, since the additional terms  we find for reduced supersymmetry do not have any new pole  singularity in such kinematic limits. 

\item[{\bf 3.}]
Inspecting Table  \ref{tab:difference-remainders} we can further infer that the difference between the remainders of $\O_{\mathcal  S}$ and  $\O_{\mathcal  C}$ when computed in $\N\!=\!2$ SYM only contains terms of  transcendentality degree 1 and 0. 

\item[{\bf 4.}]
A final  comment is in order. Throughout this paper we have used four-dimensional amplitudes and form factors as inputs to the unitarity cuts. Consequently 
our integrands might miss so-called ``$\mu^2$-terms", which might survive loop integration and could affect some of the rational numbers quoted in Table \ref{tab:difference-remainders} (see \cite{Johansson:2017bfl,Kalin:2018thp} for recent examples of the appearance of such terms in $\mathcal{N}\!=\!2$ SQCD).
\end{itemize}

\subsection{$\N\!=\!1$ SYM}
\subsubsection{The $\N\!=\!1$ SYM form factors}

For $\N\!=\!1$ SYM, the operators $\O_{\mathcal  S}$ and $\O_{\mathcal  C}$ have the same (non-minimal) tree-level form factors and as such their remainders are identical. As a result, the integrand for the two-loop form factor of $\O_{\mathcal S}$, $\O_{ \mathcal{C}}$ computed in $\N\!=\!1$ SYM can be expressed in terms of its difference with respect to the $\N\!=\!4$ SYM result for $\O_\mathcal{C}$, as
\begin{align}
F^{(2)}_{\N=1\,\O_{\mathcal S},\O_{ \mathcal{C}}}\,=\,F^{(2)}_{\N=4\, \O_{\mathcal  C}}+\Delta_{\N=1}\,,\qquad \Delta_{\N=1}\,=\, \sum_{i=5}^{15}  N''_i \times I_i\,,
\end{align}
with the numerators listed in \eqref{eq:ff-emeryN1N4}.

\subsubsection{The $\N\!=\!1$ SYM remainders}
Similarly to the  $\N\!=\!2$ SYM case, by demanding the finiteness of the remainder function we can fix the parameters $\gamma_0$, $\rho_1$ and $\rho_2$ appearing in the  renormalised remainders, with the result:
\begin{align}\label{valuesN1}
\gamma_0\,=\,-\frac{3}{2}\,,\qquad 
 \textcolor{black}{\rho_2}
\,=\,-\frac{9}{4}\,,\qquad \rho_{1}\,=\,-\frac{9}{2}\,.
\end{align}
Next, we present our result    in terms of  the difference between the remainder computed in $\N\!=\!1$ SYM and those computed in $\N\!=\!4$ SYM, see Table \ref{tab:difference-remainders2}.  
\bgroup
\def\arraystretch{1.5}
\begin{table}[h]
	\centering
	\begin{tabular}{c|c|c|}\cline{1-3}
	\hline
		\multicolumn{1}{|c|}{\bf Transc.} & $\cR^{(2)}_{\N=1\,\O_{\mathcal S},\O_{ \mathcal{C}}}-\cR^{(2)}_{\N=4\,\O_{\mathcal S}}$ & $\cR^{(2)}_{\N=1\,\O_{\mathcal S},\O_{ \mathcal{C}}}-\cR^{(2)}_{\N=4\,\O_{\mathcal C}}$  \\\hline
		\multicolumn{1}{|c|}{4}&0&0\\
		\multicolumn{1}{|c|}{3}&$-\frac{15}{4}\,\zeta_2[\log(uvw)+3\log(-q^2)]-\frac{33}{4}\zeta_3$&$-\frac{15}{4}\,\zeta_2\log(uvw)-\frac{33}{4}\zeta_3$\\
		\multicolumn{1}{|c|}{2}&$\frac{243}{8}\zeta_2$&$\frac{243}{8}\zeta_2$\\
		\multicolumn{1}{|c|}{1}&$\frac{13}{2}[\log(uvw)+3\log(-q^2)]$&$\frac{9}{2}[\log(uvw)+3\log(-q^2)]$\\
		\multicolumn{1}{|c|}{0}&$-\frac{339}{8}$&$-\frac{135}{8}$\\\hline
	\end{tabular}
	\caption{\it Difference between two-loop Catani remainders of operators $\O_{\mathcal  S}$ and $\O_{\mathcal  C}$ when calculated in $\N\!=\!4$ and $\N\!=\!1$ SYM, split by transcendentality degree. }
	\label{tab:difference-remainders2}
\end{table}
\egroup

Inspecting Table \ref{tab:difference-remainders2}, we realise that the discussion in Section \ref{Sec:N=2rem} can be repeated almost  verbatim.\footnote{Including the potential
modifications to the rational numbers in Table \ref{tab:difference-remainders2} due to the omission of $\mu^2$-terms.}
The transcendentality-four part of  the  $\N\!=\!1$  remainder is identical to that in the $\N\!=\!4$ SYM theory, confirming its  universality  \cite{Brandhuber:2017bkg}.  The difference between the remainders of operators  is limited only to a small number of pure terms, {\it i.e.} terms without rational prefactors of the type $u/v$ or $u^2/v^2$ (and permutations thereof), with all the 
non-pure terms in the $\cN\!=\!1$ SYM remainder being the same  as in $\cN\!=\!4$ and  $\cN\!=\!2$ SYM, given in \eqref{eq:non-pure-3} and \eqref{eq:non-pure-2}. Only pure logarithms, and $\zeta_2$ and $\zeta_3$ terms make an appearance in the difference, without  rational prefactors. 
Again, this is consistent with the absence of  unphysical soft/collinear singularities in the remainder function, as discussed in Section \ref{Sec:N=2rem}. 

\section{Discussion}
\label{Sec:Discussion}

We conclude with a discussion of our results for the remainder functions of the operators $\OS$ and $\OC$ in the various supersymmetric theories, and of some consistency checks of our calculations. 

\begin{itemize}

\item[{\bf 1.}] The striking property of our result for the remainders in $\N\!=\!1$, $\N\!=\!2$ and pure Yang-Mills is that their  transcendentality-four part is universal and equal to that in the $\N\!=\!4$ SYM theory \cite{Brandhuber:2017bkg}.  
The difference between the remainders of operators  is restricted to   pure terms -- terms without rational prefactors of the type $u/v$ or $u^2/v^2$ (and permutations). 
Such differences for the   $\cN\!=\!2$ SYM  and $\cN\!=\!1$ SYM remainders are listed in Tables \ref{tab:difference-remainders} and \ref{tab:difference-remainders2}. 
Terms which are allowed in the difference are logarithms,  $\zeta_2$ and $\zeta_3$ terms. 


\item[{\bf 2.}] We note that the only multi-scale integrals in our basis in Table 
\ref{tab:two-loop basis} are $I_2$, $I_3$ and $I_4$, 
and these are all determined by the three-particle cut in Figure \ref{fig:allcuts}~($iii$).  Since this cut is theory and operator independent, it follows that differences between theories and operators are confined to  single-scale integrals, which can only produce logarithmic terms. This partially explains the structural similarities between remainders in different theories and with different operators.

 \item[{\bf 3.}] \textcolor{black}{The similarity between the remainders in the $\cN\!=\!2$,  $\cN\!=\!1$ and  $\cN\!=\!4$ SYM theories must have  a  reflection  in  their behaviour  under soft/collinear limits for consistency with factorisation theorems, as we now discuss. 
 In Section 6 of \cite{Part1} it was shown that the two-loop form factors of $\cO_{\cS}$ and $\cO_{\cC}$  in $\cN\!=\!4$ SYM factorise onto a subminimal form factor with two positive-helicity gluons, and importantly this quantity is theory independent. 
This can be seen by looking at the only contributing cut,  shown  in Figure~\ref{fig:subminimal} for convenience (the full calculation is presented in Section~4.7 of \cite{Part1}, to which we refer the reader for further details). 
Hence, the expectation is that soft and collinear factorisation for this particular form factor is independent of the theory and choice of operator.  In order to confirm this from our calculation, we recall that 
 the differences between  remainder functions  in different theories or for different operators  is confined to logarithmic terms and numerical constants, {\it i.e.}~the non-pure part of the two-loop remainder is universal and identical to that of   $\cN\!=\!4$ SYM (shown  for convenience in Appendix~\ref{Sec:N=4recap}). 
These differences cannot develop any additional soft/collinear singularities, thereby satisfying  the same factorisation properties as in the maximally supersymmetric theory. This is an important consistency check of our  results.
 \begin{figure}[h]
\centering
\includegraphics[width=0.3\textwidth]{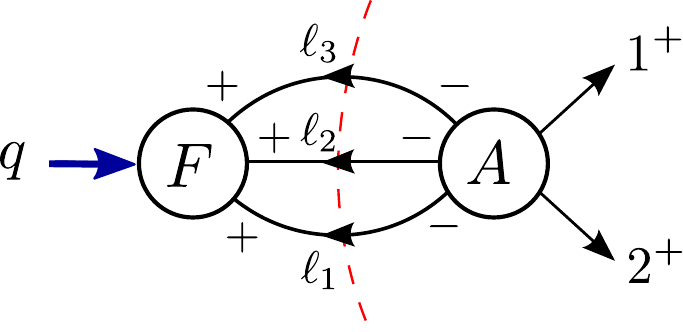}
\caption{\it Triple cut  of the two-loop subminimal form factor $F^{(2)}_{\O_{\mathcal S},\cO_\cC}(1^+,2^+;q)$. This cut is independent of the theory and  the operator chosen, because the three cut legs can only be gluons.}
\label{fig:subminimal}
\end{figure}
}

\item[{\bf 4.}]  An additional consistency check on our result can be performed by computing the values of the parameters $\gamma_0$ and $\rho_2$ entering the Catani remainder \eqref{eq:Catani} through the renormalised form factors. In our calculation these parameters  can be determined by requiring the finiteness of the remainder. 
To this end,  we consider the beta function for the operator coupling $\lambda$ introduced in \eqref{eq:lambdaren}. Since the left-hand side of that  expression is independent of $\mu$, the following renormalisation group equation must hold: 
\begin{align}
\label{rge}
0\,&=\,\mu \frac{\partial}{\partial \mu} \left\{\lambda(\mu)\left[1-a(\mu)\frac{\gamma_0}{\eps}+\frac{a(\mu)^2}{2}\left(\frac{
 \textcolor{black}{\rho_2}
}{\eps^2}-\frac{\rho_1}{\eps}\right)\right]+\O(a(\mu)^3)\right\}\, . 
\end{align}
Defining $\gamma_\lambda$ through
\begin{align}
\mu\frac{\partial \lambda(\mu)}{\partial \mu}  \,\coloneqq \,   \lambda (\mu) \ \gamma_\lambda
\, , 
\end{align}
we find that \eqref{rge} leads to the two relations
\begin{align}
\gamma_\lambda\,=\, -2\,a(\mu)\, \Big[\gamma_0 \, + \, a(\mu)\rho_1\Big]\,,
\end{align}
and 
\begin{align}\label{eq:consistency}
\gamma_0^2\, + \, \beta_0\gamma_0 \ = \ 
 \textcolor{black}{\rho_2}
\, . 
\end{align}
Here \eqref{eq:consistency} follows from  demanding  the cancellation of the $\epsilon^{-1}$ poles in the expression for 
$\gamma_\lambda$ and is a general relation that must be obeyed  by the one-loop quantities $\beta_0$ and $\gamma_0$ and the two-loop quantity $\rho_2$. The values we have determined, quoted  in \eqref{valuesN2}  and \eqref{valuesN1} for $\N\!=\!2$ and $\N\!=\!1$ SYM, respectively, obey \eqref{eq:consistency}, thereby providing a strong consistency check of our result.

\item[{\bf 5.}] Next we  comment that our calculation has independently confirmed for $\cN\!=\!1,2,4$ SYM the values for $K$ and $H^{(2)}$ which enter the two-loop Catani remainder \eqref{eq:Catani} obtained in \cite{Bern:2002tk,Bern:2003ck}, 
see  {\it e.g.}~\eqref{eq:K} and \eqref{eq:H} of \cite{Bern:2004cz}.
The  particular values of these constants are crucial to ensure the infrared  finiteness of  the renormalised remainder.

 \item[{\bf 6.}] 
 The constant  $\rho_1$ is the two-loop anomalous dimension of the operators considered here (divided by $-2$) provided that the $\mu^2$-terms do not alter the $\cO (1/\eps)$ part of our result (note  that we have used four-dimensional generalised unitarity throughout).  
Similar calculations  making use of   four-dimensional cuts done  in $\cN\!=\!4$ SYM  \cite{Brandhuber:2016fni,Brandhuber:2017bkg} led to the correct Konishi anomalous dimension  in that theory \cite{Bianchi:2000hn}. It would be interesting to check the values of $\rho_1$ (and  the corresponding anomalous dimensions) determined in this paper with an independent calculation.

\end{itemize}
 
The beautiful simplicity of our results for any amount of supersymmetry  clearly calls for a deeper explanation going beyond brute-force perturbative calculations. We will  come back to this in future work.


\section*{Acknowledgements}
	
We would like to thank Zvi Bern, John Joseph Carrasco  and Henrik Johansson for very helpful discussions, Claude Duhr for sharing a package for handling polylogarithms, and Lance Dixon, Claude Duhr,  Paul Heslop and Elli Pomoni  for stimulating conversations. 
The work of AB and GT was supported by the Science and Technology Facilities Council (STFC) Consolidated Grant ST/L000415/1  
\textit{``String theory, gauge theory \& duality"}. The work of MK is supported by an STFC quota studentship. BP is funded by the ERC Starting Grant 637019 ``\emph{MathAm}''. AB and GT would like to thank the KITP at the University of California, Santa Barbara, where their research was supported by the National Science Foundation under Grant No.~NSF PHY-1125915.
GT is grateful to the Alexander von Humboldt Foundation for support through a Friedrich Wilhelm Bessel Research Award, and to the Institute for Physics and IRIS Adlershof at Humboldt University, Berlin, for their warm hospitality. AB and GT  thank the organisers 
of the 2017 IFT Christmas workshop and the 2018 Bethe forum  \cite{talk12}, where  the results of this work were anticipated.  

\newpage	
\appendix

\section{One-loop integral functions}\label{App:Integrals} 
 
Throughout the paper, we use the following conventions for the one-loop massless scalar integrals in dimensional regularisation (upper/lower-case letters correspond to massive/massless momenta) \cite{Bern:1994cg}:
\begin{table}[h!]
\centering
\begin{tabular}{cl}
$\pic{0.6}{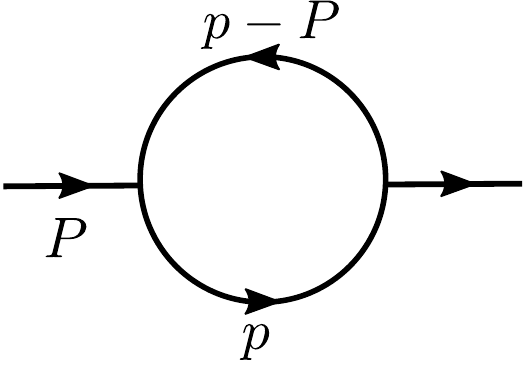}$& $ \,=\, \displaystyle\int  \dfrac{d^{4-2\epsilon}p}{(2\pi)^{4-2\epsilon}}\dfrac{1}{p^2(p-P)^2}  \,=\, i\, \frac{c_{\Gamma}}{\epsilon(1-2\epsilon)}\left(-{P^2\over \mu^2}\right)^{-\epsilon}\ ,$\\[10pt]
$\pic{0.7}{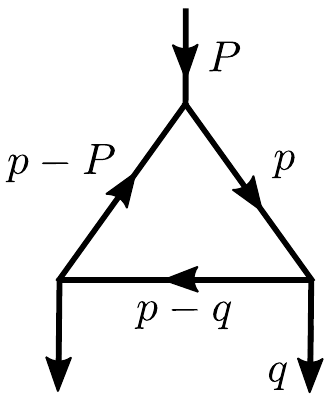}$ &$\,=\, \displaystyle \int \dfrac{d^{4-2\epsilon}p}{(2\pi)^{4-2\epsilon}}\dfrac{1}{p^2(p-q)^2(p-P)^2}= -i\, \frac{c_\Gamma}{\epsilon^2}
\dfrac{\ \ \left(-P^2 / \mu^2 \right)^{-\epsilon}}{(- P^2)}\ ,$ \\[10pt]
$\pic{0.7}{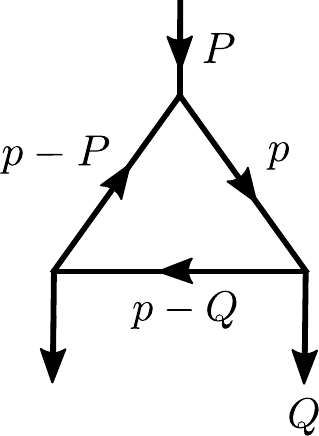}$ &$\,=\, \displaystyle \int \dfrac{d^{4-2\epsilon}p}{(2\pi)^{4-2\epsilon}}\dfrac{1}{p^2(p-Q)^2(p-P)^2}\,=\, -i\,\frac{c_\Gamma}{\epsilon^2}\frac{(-P^2/ \mu^2)^{-\epsilon}-(-Q^2/\mu^2)^{-\epsilon}}{(-P^2)-(-Q^2)}\ ,$
\\[10pt]
$\pic{0.8}{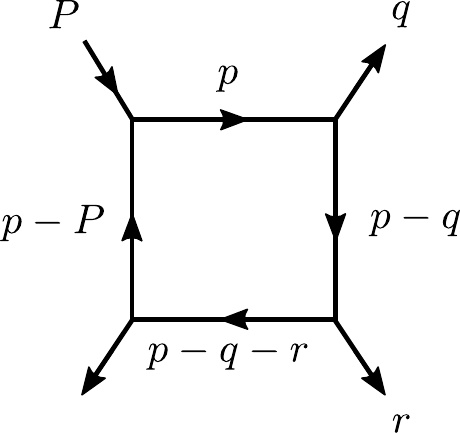} $ & $
\,=\, \displaystyle \int\! \dfrac{d^{4-2\eps} p}{(2\pi)^{4-2\eps}}\,\dfrac{1}{p^2 (p-q)^2(p-q-r)^2(p-P)^2}$ \\
& $ \displaystyle \,=\, -i\,\frac{2c_{\Gamma}}{st}\Big\{-\frac{1}{\eps^2}\Big[\Big(-{s\over \mu^2}\Big)^{-\eps}+\Big(-{t\over \mu^2}\Big)^{-\eps} -\Big(-{P^2\over \mu^2}\Big)^{-\eps}\Big] $
\\[15pt]
& $\displaystyle \,+ \text{Li}_2\Big(1-\frac{P^2}{s}\Big) + \text{Li}_2\Big(1-\frac{P^2}{t}\Big) +\frac{1}{2}\log^2\Big(\frac{s}{t}\Big) + \frac{\pi^2}{6}\Big\}\ .$
\end{tabular}
\end{table}

\noindent 
where
\begin{align*}
c_{\Gamma} \,=\, \frac{1}{(4\pi)^{2-\epsilon}} \frac{\Gamma(1+\epsilon)\Gamma(1-\epsilon)^2}{\Gamma(1-2\epsilon)}\,.
\end{align*}
	
\newpage

\section{Numerators}\label{App:Numerators}

In this appendix we present the numerators of the integral topologies which constitute the two-loop integrands for minimal form factors of $\O_{\mathcal S}$ and $\O_{\mathcal C}$ in $\N\!=\!2,1$ SYM. The integral topologies, denoted as $I_i$, $i=1,\ldots,15$ are presented in Table \ref{tab:two-loop basis}.

\subsection{Two-loop integrand for  the $\cO_\cS$ form factor in $\N\!=\!2$ SYM}
\label{app:susyN2}
The integrand for the two-loop form factor of $\O_{\mathcal  S}$ computed in $\N\!=\!2$ SYM can be expressed in terms of its difference with respect to the $\N\!=\!4$ SYM result presented in Appendix B.1 of \cite{Part1}, as
\begin{align}\nonumber
F^{(2)}_{\N=2\,\O_{\mathcal  S}}\,=\,F^{(2)}_{\N=4\, \O_{\mathcal  S}}+\Delta_{\N=2\,\O_\mathcal{S}}\,,\qquad \Delta_{\N=2\,\O_\mathcal{S}}\,=\, \sum_{i=5}^{15} N'_i \times I_i\,,
\end{align}
with the numerators
\begin{align}\label{eq:ff-emeryN2N4}
\begin{split}
N'_5\,&=\,\frac{2 s_{3 k} s_{2\ell}}{3 s_{23}}-\frac{s_{1k} s_{2\ell}}{s_{12}}+\frac{5 s_{3 k}}{3}-\frac{4 s_{23} s_{1k}}{3 s_{12}}-\frac{s_{1k} s_{3 k}}{3 s_{12}}+\frac{s_{2\ell}^2}{3 s_{23}}+\frac{2 s_{23}}{3}+(p_2\leftrightarrow p_3,k\leftrightarrow \ell)\,,\\[5pt]
N'_6\,&=\,\frac{s_{2 k} s_{1\ell}+s_{12} s_{2 k}+s_{12} s_{3 k}-s_{23} s_{1k}}{3 s_{13}}-\frac{s_{3 k} s_{1\ell}}{3 s_{12}}+\frac{s_{2 k}+s_{3 k}}{3}-\frac{s_{23} s_{1\ell}}{s_{12}}\,,\\[5pt]
N'_7\,&=\,N'_6\,\Big|_{p_2\leftrightarrow p_3}\,, \\[5pt]
N'_8\,&=\,3-\frac{s_{1\ell}}{3s_{12}}+\frac{s_{1\ell}}{s_{13}}+\frac{4s_{12}}{3s_{13}}+\frac{2s_{2k}+s_{3k}+4s_{3\ell}}{3s_{23}}\,, \\[5pt]
N'_9\,&=\, N'_8\,\Big|_{p_2\leftrightarrow p_3}\,, \\[5pt]
N'_{10}\,&=\,1+\frac{2 (s_{2 k}+ s_{3 k})}{3 s_{23}}+\frac{s_{12} s_{2 k}+s_{12} s_{3 k}}{3 s_{13} s_{23}}+\frac{s_{13} s_{2 k}+s_{13} s_{3 k}}{3 s_{12} s_{23}}-\frac{s_{1k}+3s_{13}}{3 s_{12}}-\frac{s_{1k}+3s_{1\ell}}{3 s_{13}}\,, \\[5pt]
N'_{11}\,&=\, N'_{10}\,\Big|_{p_2\leftrightarrow p_3}\,, \\[5pt]
N'_{12}\,&=\, \frac{2}{s_{23}}+\frac{4s_{12}}{3s_{13}s_{23}}+(p_2\leftrightarrow p_3,k\leftrightarrow \ell)\,, \\[5pt]
N'_{13}\,&=\, s_{2\ell}+\frac{(s_{1k}+s_{13})s_{2\ell}-(s_{2k}+s_{23})s_{1\ell}}{s_{12}}-\frac{s_{1\ell}(s_{2k}+s_{23})}{s_{13}}\,, \\[5pt]
N'_{14}\,&=\, N'_{13}\,\Big|_{p_2\leftrightarrow p_3}\,,\\[5pt]
N'_{15}\,&=\,2\,\frac{\Tr_+(1\ell k132)}{s_{12}s_{13}}\,.
\end{split}
\end{align}
\subsection{Two-loop integrand for  the $\cO_\cC$ form factor in $\N\!=\!2$ SYM}
\label{app:compN2}
The integrand for the two-loop form factor of $\O_{\mathcal  C}$ computed in $\N\!=\!2$ SYM can be expressed in terms of its difference with respect to the $\N\!=\!4$ SYM result presented in Appendix B.2 of \cite{Part1}, as
\begin{align}\nonumber
F^{(2)}_{\N=2\,\O_{\mathcal  C}}\,=\,F^{(2)}_{\N=4\, \O_{\mathcal  C}}+\Delta_{\N=2\,\O_\mathcal{C}}\,,\qquad \Delta_{\N=2\,\O_\mathcal{C}}\,=\, \sum_{i=5}^{15} \hat N_i \times I_i\,,
\end{align}
with the numerators
\begin{align}\label{eq:ff-cptN4N4}
\hat N_5\,&=\,\frac{s_{1k} s_{2 \ell}s_{3 k}}{s_{12} s_{23}}+\frac{s_{1k} s_{3 k}}{s_{12}}+(p_2\leftrightarrow p_3,k\leftrightarrow \ell)\,,\nonumber\\[5pt]
\hat N_6\,&=\,-\frac{s_{23} s_{1\ell}}{s_{12}}\,,\nonumber\\[5pt]
\hat N_7\,&=\,\hat N_6\,\Big|_{p_2\leftrightarrow p_3}\,, \nonumber\\[5pt]
\hat N_8\,&=\,\frac{2 s_{1\ell}}{s_{12}}+\frac{s_{1k}+s_{1\ell}}{s_{13}}-\frac{s_{2k}+s_{3 k}+s_{3\ell}}{s_{23}}-\frac{ s_{1\ell}s_{2 k}}{s_{12} s_{23}}-\frac{(s_{1\ell}+s_{12})s_{2k}+(s_{3k}+s_{3\ell})s_{12}}{s_{13} s_{23}}\,, \nonumber\\[5pt]
\hat N_9\,&=\, \hat N_8\,\Big|_{p_2\leftrightarrow p_3}\,, \nonumber\\[5pt]
\hat N_{10}\,&=\, 1-\frac{s_{1\ell}}{s_{13}}+\frac{s_{13}}{s_{12}}\,,\nonumber\\[5pt]
\hat N_{11}\,&=\, \hat N_{10}\,\Big|_{p_2\leftrightarrow p_3}\,, \nonumber\\[5pt]
\hat N_{12}\,&=\, \frac{s_{1\ell}}{s_{12}s_{23}}-\frac{s_{12}}{s_{13} s_{23}}-\frac{1}{s_{23}}+(p_2\leftrightarrow p_3,k\leftrightarrow \ell)\,,\nonumber\\[5pt]
\hat N_{13}\,&=\,s_{2 \ell}+\frac{s_{1k} s_{2\ell}-s_{1\ell}s_{2 k}-s_{1\ell}s_{23}+s_{13} s_{2\ell}}{s_{12}}-\frac{ s_{1\ell}\left(s_{2 k}+s_{23}\right)}{s_{13}}\,,\nonumber\\[5pt]
\hat N_{14}\,&=\, \hat N_{13}\,\Big|_{p_2\leftrightarrow p_3}\,,\nonumber\\[5pt]
\hat N_{15}\,&=\,2\,\frac{\Tr_+(1\ell k132)}{s_{12}s_{13}}\,.
\end{align}

\subsection{Two-loop integrand for  the $\cO_\cS$ and $\cO_\cC$  form factors in $\N\!=\!1$ SYM}
\label{app:compandsusyN1}
Finally,  we quote  the result for the two-loop form factors calculated in $\N\!=\!1$ SYM. As explained in Section \ref{Sec:2}, 
there is no difference between the form factors of the supersymmetric and component operators for our particular  external state. 
As a result, the integrand for the two-loop form factor of $\O_{\mathcal  S},\,\O_{\mathcal C}$ computed in $\N\!=\!1$ SYM can be expressed in terms of its difference with respect to the $\N\!=\!4$ SYM result for $\O_\mathcal{C}$ presented in Appendix B.2 of \cite{Part1}, as
\begin{align}\nonumber
F^{(2)}_{\N=1\,\O_{\mathcal S}, \O_{\mathcal C}}\,=\,F^{(2)}_{\N=4\, \O_{\mathcal  C}}+\Delta_{\N=1}\,,\qquad \Delta_{\N=1}\,=\, \sum_{i=5}^{15}  N''_i \times I_i\,,
\end{align}
with the numerators
\begin{align}
\label{eq:ff-emeryN1N4}
N''_i\,&=\,\frac{3}{2}\hat N_i,\quad i=5,\dots,15\ .
\end{align}
	\section{The  $\N\!=\!4$ SYM remainder functions}
	\label{Sec:N=4recap}
	
	In this appendix we quote the expression of the $\cN\!=\!4$ SYM remainder function computed in \cite{Brandhuber:2017bkg}. 
	In fact we will need a small modification of that  result, since in this paper we are using the Catani definition of the remainder function, while in \cite{Brandhuber:2017bkg} we used the  BDS definition (which is standard in $\cN\!=\!4$ SYM).
	The $\cN\!=\!4$ SYM Catani remainder is related to the BDS remainder as
	\begin{align}
	\cR^{(2)}_{\O,\text{Catani}}\,=\,\cR^{(2)}_{\O,\text{BDS}} -\zeta_3\,\big[ 3\log(-q^2)+\log(uvw)-6\big]-\frac{33}{8} \zeta_4 \ , \quad \O = \O_{\mathcal{S}},\O_{\mathcal{C}}\ .
	\end{align}
	Finally we quote the $\N\!=\!4$ two-loop BDS remainder of $\O_{\mathcal S}$ and $\O_{\mathcal C}$ obtained in \cite{Brandhuber:2017bkg}. At each transcendentality degree $k<4$, denoted by $\cR^{(2)}_{\O;k} $, there are pure terms and terms that are multiplied by rational prefactors that depend on the kinematics, that is
	\begin{align}
	\cR^{(2)}_{\O;k} \,=\,\cR^{(2)}_{\O;k}\Big|_{\rm pure}+\cR^{(2)}_{\O;k}\Big|_{\text{non-pure}}\ .
	\end{align}
	Explicitly we have that at transcendentality four there is only a pure term which is identical to the BPS two-loop remainder of \cite{Brandhuber:2014ica},
	\begin{align}\label{eq:remainderBPS}
	\begin{split}
	\cR^{(2)}_{\O_{\mathcal S};4}\ =\	\cR^{(2)}_{\rm BPS} \ = \ 
	& -\frac{3}{2}\, \text{Li}_4(u)+\frac{3}{4}\,\text{Li}_4\left(-\frac{u v}{w}\right) 
	-\frac{3}{2}\log(w) \, \text{Li}_3 \left(-\frac{u}{v} \right)+\frac{1}{16} {\log}^2(u)\log^2(v)  \\
	&+{\log^2 (u) \over 32} \Big[ \log^2 (u) - 4 \log(v) \log(w) \Big]+{\zeta_2 \over 8 }\log(u) \Big[ 5\log(u)- 2\log (v)\Big]  \\
	&+{\zeta_3 \over 2} \log(u) + \frac{7}{16}\, \zeta_4 + {\rm perms}\, (u,v,w) \, . 
	\end{split}
	\end{align}
	At transcendentality three, there is a pure term and a non-pure term, namely
	\begin{align}
	\begin{split}
	&\cR^{(2)}_{{\O_{\mathcal S}};3}\Big|_{\rm pure}\, =\, 
	\text{Li}_3(u)+ \text{Li}_3(1-u)- {1\over 4} \log^2(u) \log \left({v w\over (1-u)^2} \right)
	+{1\over 3} \log (u) \log (v) \log (w) \\
	&\phantom{\cR^{(2)}_{{\O_{\mathcal S}};3}\Big|_{\rm pure}\,=\,}
	+\zeta_2 \log (u) -{5\over 3}\zeta_3 + \,\text{perms}\, (u,v,w)\, ,
	\end{split}\\
	\label{eq:non-pure-3}
	\begin{split}
	&\cR^{(2)}_{{\O_{\mathcal S}};3}\Big|_{\text{non-pure}} \,=\, 	 \frac{u}{w}\Big\{\Big[-\text{Li}_3\left(-{u\over w}\right)+\log(u)\text{Li}_2\left({v\over 1-u}\right)+{1\over 2}\text{Li}_3\left(-{uv\over w}\right)+{1\over 12}\log^3(w) 
	\\
	&	\phantom{\cR^{(2)}_{{\O_{\mathcal S}};3}\Big|_{\text{non-pure}}\,=\,}-{1\over 2}\log(1-u)\log(u)\log\left({ w^2\over 1-u }\right)+{1\over 2}\log(u)\log(v)\log(w)+(u\leftrightarrow v) \Big]  \\
	&	\phantom{\cR^{(2)}_{{\O_{\mathcal S}};3}\Big|_{\text{non-pure}}\,=\,}+\text{Li}_3(1-v)-\text{Li}_3(u)+{1\over 2}\log^2(v)\log\left({1-v\over u}\right)
	\\
	&	\phantom{\cR^{(2)}_{{\O_{\mathcal S}};3}\Big|_{\text{non-pure}}\,=\,}\ 
	- \ 
	\zeta_2 \log\left( {u v\over w}\right)\Big\}+ \,\text{perms}\, (u,v,w)\, .
	\end{split}
	\end{align}
	Likewise, at transcendentality two, we have
	\begin{align}
	\begin{split}
	&\cR^{(2)}_{{\O_{\mathcal S}};2}\Big|_{\text{pure}}\,=\, -\text{Li}_2(1-u)-\log^2(u)+{1\over 2}\log(u)\log(v)-{13\over2}\zeta_2 + \,\text{perms}\, (u,v,w)\,\,\end{split}\\
	\label{eq:non-pure-2}
	\begin{split}
	&\cR^{(2)}_{{\O_{\mathcal S}};2}\Big|_{\text{non-pure}}\,=\,
	\frac{u^2}{v^2}\Big[\text{Li}_2(1 - u) + \text{Li}_2(1 - v) + \log(u)
	\log(v)-\zeta_2\Big]\,+ \,\text{perms}\, (u,v,w)\ .
	\end{split}
	\end{align}
	Finally, the transcendentality one and zero are simply
	\begin{align}
	\cR^{(2)}_{{\O_{\mathcal S}};1}\,=\, &\left(-4 + \frac{v }{w}+ \frac{u^2 }{2 v w} \right) \log(u)\, + \,\text{perms}\, (u,v,w)\, ,\\[10pt]
	\cR^{(2)}_{{\O_{\mathcal S}};0}\,=\, &7 \left(12+\frac{1}{uvw} \right)\,.
	\end{align}
	For $\O_{\mathcal{C}}$ we have
	\begin{align}
	\begin{split}
	\cR^{(2)}_{\O_{\mathcal C};i}\,&=\,\cR^{(2)}_{\O_{\mathcal S};i}\,,\qquad i=4,3,2\,,\\
	\cR^{(2)}_{{\O_{\mathcal C}};1}\,&=\,\cR^{(2)}_{{\O_{\mathcal S}};1}+2\log(u v w) \, , 
	\\
	\cR^{(2)}_{{\O_{\mathcal C}};0}\,&=\,\cR^{(2)}_{{\O_{\mathcal S}};0}-\frac{51}{2}\,.
	\end{split}
	\end{align}

	\pagebreak
	\bibliographystyle{utphys}
	\bibliography{remainder}

\end{document}